\documentclass{article}

\def\giorno{8/11/2018}

\usepackage{color}
\usepackage{hyper ref}
\usepackage{amsmath,amsfonts,amssymb}

\def\a{\alpha}
\def\b{\beta}
\def\ga{\gamma}
\def\eps{\varepsilon}
\def\vphi{\varphi}
\def\la{\lambda}

\def\s{\sigma}

\def\vphi{\varphi}

\def\S{{\mathcal S}}

\def\De{\Delta}
\def\La{\Lambda}

\def\pa{\partial}

\def\xb{{\bf x}}

\def\o+{\oplus}

\def\ss{\subset}

\def\<{\langle}
\def\>{\rangle}

\def\({\left(}
\def\){\right)}
\def\[{\left[}
\def\]{\right]}
\def\=#1{\bar #1}
\def\~#1{\widetilde #1}
\def\wt#1{\widetilde #1}
\def\.#1{\dot #1}
\def\^#1{\widehat #1}

\def\"#1{\ddot #1}

\def\eeq{\end{equation}}
\def\beq{\begin{equation}}

\def\beql#1{\begin{equation} \label{#1}}

\def\eqref#1{(\ref{#1})}

\def\EOR{ \hfill $\odot$ \medskip}

\def\EOP{ \hfill $\triangle$ \medskip}

\def\symmref{AVL,CGbook,Olver1,Olver2,Stephani}

\def\RK{6}
\def\RKK{7}
\def\LDS{8}

\begin{document}

\title{W-Symmetries of Ito stochastic differential equations}

\author{G. Gaeta$^{1,2}$\thanks{giuseppe.gaeta@unimi.it}\\
${}^1$ {\it Dipartimento di Matematica, Universit\`a degli Studi di
Milano,} \\ {\it via Saldini 50, 20133 Milano (Italy)} \\
${}^2$ {\it SMRI, 00058 Santa Marinella (Italy)} \\
}

\date{\giorno }

\maketitle

\begin{abstract}

We discuss W-symmetries of Ito stochastic differential equations,
introduced in a recent paper by Gaeta and Spadaro [J. Math. Phys.
2017]. In particular, we discuss the general form of acceptable
generators for continuous (Lie-point) W-symmetry, arguing they are
related to the (linear) conformal group, and how W-symmetries can
be used in the integration of Ito stochastic equations along
Kozlov theory for standard (deterministic or random) symmetries.
It turns out this requires, in general, to consider more general
classes of stochastic equations than just Ito ones.

\end{abstract}

\section{Introduction}

In a recent paper \cite{GS17} we have discussed in general terms
symmetry of (systems of) Stochastic Differential Equations in Ito
form, \beql{eq:Ito} d x^i \ = \ f^i (x,t) \, dt \ + \ \s^i_{\ k}
(x,t) \, d w^k \ . \eeq (Here and below sum over dummy indices is
routinely understood.) We have argued that albeit apparently one
could consider general vector fields in $(x,t;w)$ space, i.e.
\beql{eq:Xgen} X \ = \ \vphi^i (x,t;w) \, \frac{\pa}{\pa x^i} \ +
\ \tau (x,t;w) \, \frac{\pa}{\pa t} \ + \ h^k (x,t;w) \,
\frac{\pa}{\pa w^k} \ , \eeq some limitations are actually in
order on the functions $\vphi , \tau , h$, see below.

Based on these, we have proposed a classification of different
types of symmetries, and investigated the determining equations
characterizing them for a given Ito equation \eqref{eq:Ito}.

In later work \cite{GL1,GL2}, after clarifying how the relevant
symmetries of an Ito SDE are preserved under a change of variables
despite the non-geometric transformation properties of Ito
equations under these \cite{GL1}, our discussion and
classification were useful in order to extend the Kozlov theory
\cite{Koz1,Koz2,Koz3,Koz2018} relating symmetry and -- complete or
partial -- integrability of SDEs. In particular, it was shown that
the sufficient conditions identified by Kozlov for this in the
case (according to our classification, see below) of
\emph{deterministic} symmetries, are also necessary; and the
theory was also extended to \emph{random} symmetries \cite{GL2},
albeit in this case we only treated scalar equations (we will fill
this gap by treating the case of systems, in
sect.\ref{sec:ransyst} below)\footnote{It should be mentioned that
in recent work \cite{Koz18a}, Kozlov considered the situation
where the Ito equation admits a conserved quantity, and studied
the consequence of this on the symmetries and their algebraic
structure; it was also shown how, even in this case, there is a
correspondence between the symmetries of the Ito and of the
associated Stratonovich equation \cite{Koz18b}, confirming the
results of \cite{GL1}.}

The purpose of the present work is to discuss the extension of
this approach and the results mentioned above to the other case
allowed by our  classification, i.e. to \emph{W-symmetries}. These
are symmetries directly acting -- beside the $x^i$ and $t$
variables -- on the Wiener process $w^i$ as well \cite{GS17}.

We will also denote vector fields and symmetries \emph{not} acting
on (but possibly depending on) the $w^i$ variables, as
\emph{standard} ones, for ease of reference; thus standard
symmetries comprise both deterministic and random ones.

\bigskip\noindent
Let us now briefly sketch the \emph{plan of the paper}. We will
start by recalling, in Section \ref{sec:symmIto}, our discussion
about the limitations to be put on \eqref{eq:Xgen} to get
admissible symmetries, and hence our classification for the three
types of admissible symmetries (deterministic, random, and
W-symmetries) together with the relevant concept of \emph{simple}
symmetry. We will also briefly recall, in Section \ref{sec:geom},
how it is possible to use (deterministic or random) symmetries of
Ito SDE despite the transformation properties of these.

We will then discuss, in Section \ref{sec:Koz}, Kozlov theory for
standard symmetries; in particular we will recall how the presence
of simple symmetries -- deterministic or random -- allow to
integrate a scalar SDE. In Section \ref{sec:Kozsyst} we will
consider systems; in particular, in \ref{sec:Kozsystdet} we recall
how one can use deterministic symmetries (provided a Lie algebraic
condition, analogous to the one met when dealing with
deterministic equations, is satisfied) to partially integrate,
i.e. reduce to a smaller dimension, a system of Ito equations. As
mentioned above, our previous work only considered symmetry
reduction (actually, in this case, integration) under random
symmetries for the case of scalar equations; in Section
\ref{sec:ransyst} we will extend that discussion to the general
case, i.e. systems of Ito equations (this result is new, but is a
straightforward extension of those already present in the
literature).

We will then be ready to introduce the most relevant -- and
original -- part of our work, namely the extension of the theory
developed so far for deterministic or random symmetries to the
third case in our classification, i.e. for W-symmetries.

This will first of all require again to discuss more precisely
what kind of transformation could and should be considered, which
is the subject of Section \ref{sec:Wsymm}. After this, we will
have to extend the discussion of Section \ref{sec:geom} to the
case of W-symmetries; this will be done in Section \ref{sec:WSIS},
and we will find that the extension is not complete.

After this we will finally be able to tackle the extension of
Kozlov theory to W-symmetries, in Section \ref{sec:Wkoz}. Again we
will find that the extension is not complete; in particular we
will see that albeit W-symmetries are of help in educing or
integrating Stochastic Differential Equations, this will in
general go though mapping an Ito equation into a more general type
of stochastic equation. This also means that the existing results
about multiple symmetry reduction cannot be applied in the case of
multiple W-symmetries.

In the final Section \ref{sec:discuss} we will summarize and
discuss our findings.

We also have two Appendices, devoted to the (simpler) special case
of scalar equations. In Appendix A we derive, in the simplified
one-dimensional setting, our basic result about the correspondence
of W-symmetries for an Ito and the associated Stratonovich
equations; in Appendix B we show that not all vector fields can be
realized as nontrivial W-symmetries of stochastic equations,
discussing in detail some one-dimensional examples.

The symbol $\odot$ will mark the end of a Remark or of an Example.

\subsection*{Acknowledgements}

I thank C. Lunini, L. Peliti and F. Spadaro for interesting
discussion on symmetries of SDEs in general, and on this research
in particular; the communication by prof. R. Kozlov of a simple
but significant Example (see Example 2 below) was also very useful
to focus my ideas. A substantial part of this work was performed
while I was in residence at SMRI over the summer 2018.


\section{Standard symmetries of Ito equations}
\label{sec:symmIto}

When we consider an Ito equation\footnote{By this we always mean
possibly a system, unless otherwise specified. Note that -- albeit
this will in general not be used -- we can also assume $\s$ to be
non-degenerate or, passing to suitable coordinates, we would have
an Ito system coupled to deterministic ODEs.} \eqref{eq:Ito},
applying the vector field \eqref{eq:Xgen} produces a map \beql{eq:Xgenmap} x^i
\to \wt{x}^i = x^i + \eps \vphi^i (x,t;w) \ , \ \ t \to \wt{t} =t
+ \eps \tau (x,t;w) \ , \ \ w^i \to \wt{w}^k = w^k + \eps h^k
(x,t;w) \ ; \eeq this in turn maps the Ito SDE \eqref{eq:Ito} into
a, generally different, SDE.

The point is that for general choices of $\vphi^i,\tau,h^k$ the
new SDE is not even of Ito type, as discussed in detail in
\cite{GS17}. In order to ensure we remain within the framework of
Ito equations, we should introduce several limitation on these
coefficients\footnote{We stress they are always supposed to be
$C^\infty$ (we will also say just \emph{smooth}) functions of
their argument; this guarantees we are dealing with proper --
albeit possibly formal -- vector fields in the $(x,t;w)$ space.};
in particular, leaving aside for a moment the coefficients $h^k$
and hence the possibility to consider W-symmetries:

\begin{itemize}

\item The functions $\vphi^i$ are unrestricted, beside the
requirement to be smooth functions of their arguments.

\item The function $\tau$ should (be smooth and) depend
\emph{only} on $t$, with moreover $\tau' (t) > 0$ (this guarantees
the new variable $\wt{t}$ still represents time, albeit a rescaled
one).

\end{itemize}

We will from now on always assume that these restrictions on
$\tau$ are satisfied; we refer to these vector fields, and
possibly symmetries, as the \emph{admissible} ones.

\medskip\noindent
{\bf Remark 1.} Note that if $\tau \not= 0$, the rescaling of time
will affect the Wiener processes $w^i$. More precisely, their
expression $w^i = w^i (\wt{t})$ in terms of the new time $\wt{t}$
will differ from their expression $w^i = w^i (t)$ in terms of the
pristine time variable. However, this difference amounts to a
scalar factor, which is then absorbed in the coefficients $\s^i_{\
k}$ of the Ito equation; see \cite{GRQ1,GS17}.

More precisely, in this case we get $w^i
(t) \to \wt{w}^i (\wt{t} ) $ with \beq \wt{w}^i (\wt{t} ) \ = \
\sqrt{1 + \eps \tau' (t ) } \ w^i (\wt{t} ) \ ; \eeq all in all,
this amounts to the map \beql{eq:wtaumap} d w^k \ \to \ d w^k \ +
\ \eps \, \frac12 \, \( \frac{d \tau}{d t} \) \, d w^k \ := \ d
w^k \ + \ \eps \ \delta w^k \ . \eeq See e.g. \cite{GS17}, sect.
IIB, for details. \EOR
\bigskip

With these limitations, and still keeping $h^k = 0$, we have a
simple classification of maps and hence of possible symmetries.

\begin{itemize}

\item If the $\vphi^i$ do not depend on the $w^k$ variables, then
we speak of \emph{deterministic} vector fields; if they also
effectively depend on the $w^k$, we speak of \emph{random} vector
field. Note that by assumption $\tau$ only depends on $t$, if it
is present.

\item If $\tau = 0$, we speak of \emph{simple} vector fields; if
$\tau \not= 0$ (but $\tau = \tau (t)$, $\tau' (t) > 0$) we speak
of \emph{general} vector fields.
\end{itemize}

\medskip\noindent
{\bf Remark 2.} We anticipate that the Kozlov theory relating
symmetry of SDE to the possibility of reducing, and possibly
completely integrating, them makes only use of \emph{simple}
symmetries (see however \cite{Koz18a,Koz18b}); hence the special
interest of this seemingly restricted class. \EOR

\medskip\noindent
{\bf Remark 3.} As well known, when dealing with
\emph{deterministic} differential equations, there is no such
difference between symmetries acting on the time and on the
spatial variables (and actually we can consider symmetries mixing
time and the space variables). The reason for their different
standing in the present context is readily understood: in fact,
now $t$ is in all cases a smooth variable, while $x$ is a smooth
variable as the spatial coordinate, but becomes a stochastic
process when we look at solutions to the Ito equation
\eqref{eq:Ito}; thus $t$ and the $x^i$ are inherently different,
and it is no surprise that time should not be mixed with space
variables, and that the presence of symmetries acting on them will
have different consequences. \EOR


\section{Standard symmetry and change of variables}
\label{sec:geom}

The possibility of using symmetries to solve or reduce
\emph{deterministic} equations rests ultimately on the fact that
symmetries are preserved under changes of variables. This in turn
follows immediately from the fact that symmetry vector fields are
geometrical objects, and the same holds for the \emph{solution
manifold} $S_\De \ss J^n M$ representing a differential equation (or
system) $\De$ of order $n$ in the suitable Jet space
\cite{\symmref}.

It is not at all obvious that the same holds for Ito equations: as
well known, they do not transform geometrically (i.e. under the
chain rule), but in their own way -- in fact, under the Ito rule.

This point was raised and solved in some recent work \cite{GL1}.
The approach followed there was to use the Stratonovich equation
associated to a given Ito one; this transforms geometrically (this
is its main advantage, together with the related time-inversion
properties), so its symmetries are surely preserved under changes
of variables. The determining equations for an Ito equation and
for the associated Stratonovich one are different and give
different solutions \cite{GS17,Unal}; but it is known that they
have the same solutions if we restrict to either simple
(deterministic or random) symmetries, or to general symmetries
\eqref{eq:Xgen} with $\tau$ satisfying a certain third order
compatibility condition identified by Unal \cite{Unal}; this is
automatically satisfied if $\tau$ only depends on $t$, i.e. for
\emph{admissible} symmetries according to our classification
\cite{GS17} recalled above.

In other words, we have the following result \cite{GL1}; here
``simple'' refers again (as for symmetries) to the fact the $t$
variable is unaffected; we will similarly denote as ``simple
maps'' those not acting on $t$.

\medskip\noindent
{\bf Proposition 1.} {\it Admissible standard symmetries of an Ito
equation \eqref{eq:Ito} are preserved under (simple,
deterministic) smooth changes of variables $x^i =\Phi^i (y,t)$.}

\section{Standard symmetry and integrability of scalar Ito equations}
\label{sec:Koz}

With the result of the previous Section, we can start discussing
symmetries of an Ito equation and its use.

First if all we note that -- as an Ito equation lacks a
geometrical interpretation -- in this context symmetry will be an
\emph{algebraic} rather than a geometrical property. That is, we
require that the map \eqref{eq:Xgenmap}, i.e. the substitution $
x^i \to x^i + \eps \vphi^i$, $t \to t + \eps \tau$, $ w^k \to w^k
+ \eps h^k$, leaves the Ito equation \eqref{eq:Ito} invariant
\emph{at first order in $\eps$}.

In the case of interest here, i.e. disregarding for the moment
W-symmetries, and focusing on (deterministic or random)
\emph{simple} symmetries \beq X \ = \ \vphi^i (x,t:w) \ \pa/\pa x^i \ ,
\eeq it can be proven \cite{GS17} that they comply with
the \emph{determining equations} (for simple symmetries)
\begin{eqnarray}
\pa_t \vphi^i \ + \ f^j \, \pa_j \vphi^i \ - \ \vphi^j \, \pa_j
f^i &=& - \, \frac12 \ \triangle \vphi^i \ , \\
\^\pa_k \vphi^i \ + \ \s^j_{\ k} \, \pa_j \vphi^i \ - \ \vphi^j \,
\pa_j \s^i_{\ k} &=& 0 \ ; \end{eqnarray} here we have used the
notation \beql{eq:not} \pa_t \ := \ \pa / \pa t \ , \ \ \pa_i \ :=
\ \pa / \pa x^i \ , \ \ \^\pa_k \ := \ \pa / \pa w^k \ ; \eeq and
the symbol $\triangle$ denotes the \emph{Ito Laplacian}
\beql{eq:ItoLapl} \triangle u \ := \ \sum_{k=1}^n \ \frac{\pa^2
u}{\pa w^k \pa w^k} \ + \ \sum_{j,k=1}^n \ \left( \s \, \s^T
\right)^{jk} \ \frac{\pa^2 u}{\pa x^j \, \pa x^k} \ + \ 2 \
\sum_{j,k=1}^n \s^{jk} \ \frac{\pa^2 u}{\pa x^j \, \pa w^k} \ .
\eeq These notations will be used routinely in the following.


Let us first consider the case of a scalar equation; then the
presence of a simple symmetry guarantees that the equation can be
explicitly integrated, i.e. transformed into an Ito integral. The
result is constructive, in that the symmetry determines the
appropriate change of variables.

This result holds for any standard simple symmetry, but in the
case of random ones some additional condition should also be
checked.

\subsection{Deterministic symmetries}

We start with the case of simple deterministic symmetries. Here we
have the following result, due to Kozlov \cite{Koz1} (see also
\cite{GL2}):

\medskip\noindent
{\bf Proposition 2.} {\it The scalar SDE \beql{eq:K1} d y \ = \
\wt{f} (y,t) \ d t \ + \ \wt{\s} (y,t) \ d w \eeq can be
transformed by a simple deterministic map $y = y (x,t)$ into
\beql{eq:K2} d x \ = \ f(t) \, d t \ + \ \s (t) \, d w \ , \eeq
and hence explicitly integrated in Ito sense, \emph{if and only
if} it admits a simple deterministic symmetry.

If the generator of the latter is $ X  = \vphi (y,t) \pa_y$, then
the change of variables $y =  F (x,t)$ transforming \eqref{eq:K1}
into \eqref{eq:K2} is the inverse to the map $x = \Phi (y,t)$
identified by
$$ \Phi (y,t) \ = \ \int \frac{1}{\vphi (y,t) } \ d
y \ . $$}

\medskip\noindent
{\bf Remark 4.} We stress that here the ``only if'' refers to the
transformation \emph{by a deterministic map}. We will see in a
moment that the transformation is possible also in case there is
no deterministic symmetry but a random symmetry is present; but in
this case this is achieved by a \emph{random map} rather than a
deterministic one. See also Remark {\RK}  in this sense. \EOR

\medskip\noindent
{\bf Remark 5.} Note that \eqref{eq:K2} provides immediately the
solution in the new variable,
$$ x (t) \ = \ x (t_0) \ + \ \int_{t_0}^t f(s) \, d s \ + \
\int_{t_0}^t \s (s) \, d w(s) \ ; $$ in order to obtain the
solution in the original variable we should of course use $y =
F(x,t)$. \EOR

\medskip\noindent
{\bf Example 1.} The Ito equation \cite{GL1}
$$ d y \ = \ \[ e^{- y} \ - \ (1/2) \, e^{-2 y} \] \, d t
\ + \ e^{- y} \, d w $$
admits the vector field $ X  = e^{- y} \pa_y $ as a symmetry
generator. By the associate change of variables $$ x \ = \ \int
\frac{1}{\vphi (y)} \ d y \ = \ \int e^y \, d y \ = \ \exp[y] \ +
\ K $$ the vector field reads $X = \pa_x$, and the initial
equation reads
$$ d x \ = \ d t \ + \ d w \ ; $$
this is readily integrated.\footnote{More precisely, we get $x(t)
= c_0 + t + w(t)$, and hence $y(t) = \log [ x(t) - K]$; a suitable
choice of the arbitrary (integration) constant $K$ guarantees
existence of the solution $y(t)$ for sufficient times $t > 0$ with
probability one. In the following Examples we will not discuss the
map of solutions back into the original variables.} \EOR

\subsection{Random symmetries} \label{sec:randomfirst}

The difference between the case where deterministic symmetries are
considered and the one where the considered symmetries are random
ones, lies in that in the case of random symmetries the associated
random change of variables could change the Ito equation into a
random system of different nature. This problem accounts for the
appearance of an extra condition, absent when one is only
considering deterministic simple symmetries. Here we just give the
relevant result, referring to \cite{GL2} for a comprehensive
discussion.

\medskip\noindent
{\bf Proposition 3.} {\it Let the Ito equation \beql{eq:dyRR} d y
\ = \ F(y,t) \, d t \ + \ S (y,t) \, d w \eeq admit the simple
random vector field $ X  =  \vphi (y,t,w) \pa_y$ as Lie-point
symmetry; define $\ga (y,t,w) := \pa_w ( 1 / \vphi )$.

If the functions $F(y,t)$, $S(y,t)$ and $\ga (y,t,w)$ satisfy the
relation \beql{eq:bcomp} S \, \ga_{t} \ + \ S_t \, \ga \ = \ F \,
\ga_{w} \ + \ (1/2) \, \[ S \, \ga_{ww} \ + \ S^2 \, \ga_{yw} \] \
, \eeq then the equation \eqref{eq:dyRR} can be mapped by a simple
random change of variables into an integrable Ito equation
\beql{eq:dxRint} d x \ = \ f(t) \, d t \ + \ \s (t) \, d w \ .
\eeq

Conversely, let the Ito equation \eqref{eq:dyRR} be reducible to
the integrable form \eqref{eq:dxRint} by a simple random change of
variables $x = \Phi (y,t;w)$. Then necessarily \eqref{eq:dyRR}
admits $X = \[ \Phi_y (y,t,w) \]^{-1} \pa_y := \varphi (y,t,w)
\pa_y$ as a symmetry vector field, and \eqref{eq:bcomp} is
satisfied with $\ga = \pa_w (1/\vphi)$.}

\bigskip\noindent
{\bf Remark \RK.} As mentioned above, see Remark
4, it is possible that an equation can be integrated by a (random)
change of variables, albeit it has no deterministic simple
symmetry; in this case it should, as stated by Proposition 3, have
a random simple symmetry. \EOR

\medskip\noindent
{\bf Example 2.} A simple example of this situation is provided by
the scalar Ito equation\footnote{This example was communicated to
me by prof. Kozlov (personal communication), whom I warmly thank.}
\beql{eq:Kex} d x \ = \ e^x \, d t \ + \ d w \ . \eeq This has no
simple deterministic symmetry (it has the deterministic symmetry
$X_0 = \pa_t$, but this is not simple and hence cannot be used for
integration), but has a simple random symmetry, $ X  = \exp [x-w]
\pa_x$, which can be used for integration. In fact, the
$X$-related new variable is \beql{eq:KCV} y \ = \ \int
\frac{1}{e^{x-w}} \ d x \ = \ - \ e^{w - x} \ , \eeq and in terms
of this we have \beql{eq:Kred} d y \ = \ e^w \, d t \ , \eeq hence
$$ y(t) \ = \ y(t_0) \ + \ \int_{t_0}^t e^{w(s)} \ d s \ . $$
It may be noted that the equation also has a W-symmetry, $ X_1 =
\pa_w$; it turns out this is not of acceptable type, as discussed
in Sect.\ref{sec:Wsymm} below. \EOR

\medskip\noindent
{\bf Remark \RKK.} Note also that eq.\eqref{eq:Kred} is \emph{not}
of Ito type; correspondingly eq.\eqref{eq:bcomp}, meant now for
\eqref{eq:Kex} and $\ga$ associated to $X$, is \emph{not}
satisfied: the l.h.s. vanishes, while the r.h.s. yields $e^w [1 +
(1/2) e^{- x}]$. This notwithstanding, eq.\eqref{eq:Kred} is
readily integrated, and hence the change of variables
\eqref{eq:KCV} allows to integrate the original equation
\eqref{eq:Kex}. This suggest that our theory can be extended,
allowing for transformations to non-Ito equation and hence for
symmetries such that the compatibility condition \eqref{eq:bcomp}
is not satisfied. We will not dwell in this direction in the
present work, but we will find that this situation is rather
generic when dealing, in later Sections, with W-symmetries. \EOR

\section{Systems}
\label{sec:Kozsyst}

The scope of Proposition 2 is quite limited, in that it only
concerns \emph{scalar} equations. On the basis of what is achieved
in the case of deterministic equations, we would expect that if a
(simple) symmetry is present for a multi-dimensional system, then
the latter can be reduced to one of lower dimension -- in this
case we also say it can be ``partially integrated''. In fact, this
is the case also for SDEs, as stated by the following Proposition
4.


Needless to say, in the case of multi-dimensional systems one
could have several (simple) symmetries, and -- in principles --
multiple reduction is possible. Once again, in the case of
deterministic equations this is the case only if the symmetries
(more precisely, only for those of the symmetries which) have a
suitable algebraic structure, i.e. which span a \emph{solvable Lie
algebra} acting with regular orbits \cite{\symmref}, and one would
expect the same kind of condition is required also in the analysis
of SDEs, as indeed is the case.

\subsection{Partial integrability and multiple deterministic symmetries}
\label{sec:Kozsystdet}

It appears that only deterministic symmetries have been considered
so far in discussing systems. We will provide a discussion of
multiple reduction by random symmetries in Section
\ref{sec:ransyst}.

Again the relevant results in this direction have been obtained by
Kozlov \cite{Koz2,Koz3} (see also \cite{GL2,Lunini}). We will be
quoting from \cite{GL2}.

\medskip\noindent
{\bf Proposition 4.} {\it Suppose the system \eqref{eq:Ito} admits
an $r$-parameter \emph{solvable} Lie algebra $\mathcal{G}$ of
simple deterministic symmetries, with generators \beq
X_{(k)}=\sum_{i=1}^n \varphi_{(k)}^i(x,t) \ \frac{\pa}{\pa x_i}
\qquad (k=1,...,r) \ , \eeq acting regularly with $r$-dimensional
orbits.

Then it can be reduced to a system of $m = (n-r)$ equations, \beq
d y^i \ = \ g^i (y^1,...,y^m;t) \, d t \ + \ \s^i_{\ k}
(y^1,...,y^m;t) \, d w^k \ \ \ \ (i,k=1,...,m) \eeq and $r$
``reconstruction equations'', the solutions of which can be
obtained by quadratures from the solution of the reduced
$(n-r)$-order system.}

\bigskip\noindent {\bf Remark \LDS.} It is convenient to label the
different generators $X_{(k)}$ of the symmetry Lie algebra
$\mathcal{G}$ according to the Lie structure of this; thus the
element $\mathcal{G}^{q}$ in the derived series of $\mathcal{G}$
will be the span of $\{ X_{(q)} ,  X_{(q+1)} , ..., X_{(r)} \}$.
We recall that the derived series is defined as $\mathcal{G}^1 =
\mathcal{G}$, and $\mathcal{G}^{q+1} = \[ \mathcal{G}^q ,
\mathcal{G}^q \]$.

Then the reduction should be performed using sequentially the
symmetries $X_{(1)}$, $X_{(2)}$, ... $X_{(r)}$, i.e. respecting
the Lie algebraic structure of $\mathcal{G}$. In this way we
obtain a sequence of reduced equations $E_k$, where $E_0$ is the
original system, $E_q$ the one obtained after reduction by
$X_{(1)}, ... , X_{(q)}$, and the reduced system mentioned in
Proposition 4 coincides with $E_r$. \EOR

\medskip\noindent
{\bf Remark 9.} It follows immediately from Proposition 4 that, in
particular, for $r=n$ the general solution of the system can be
found by quadratures. Note that here, and in the statement of
Proposition 4, this means performing Ito integrals. \EOR

\medskip\noindent
{\bf Remark 10.} In Kozlov's original paper \cite{Koz2} (see
Example 4.2 in there) this result is applied to \emph{any} linear
two-dimensional system of SDEs; see there for a detailed
discussion and results. \EOR

\medskip\noindent
{\bf Remark 11.} In view of Proposition 1 (i.e. ultimately of the
coincidence of admissible symmetries for an Ito and the associated
Stratonovich systems), the proof of Proposition 4 can be obtained
following the same approach as for deterministic differential
equations, apart from the obvious difference that now quadratures
correspond to Ito integrals. An explicit proof (with details) is
provided in \cite{GL2,Koz2,Koz3,Lunini}. \EOR


\subsection{Systems and random symmetries}
\label{sec:ransyst}

The approach discussed above, i.e. Kozlov theory, is based on
performing changes of variables related to (simple) symmetries of
the Ito equation under study. If these symmetries are random ones,
we will have to consider (simple) random changes of variables;
this introduces an additional problem, as we are then not
guaranteed to remain within the class of Ito equations (see the
discussion in Section \ref{sec:randomfirst}, in particular Remark
{\RKK}).

Let us consider a general vector Ito equation \eqref{eq:Ito}. If
we operate a general simple random change of variables, i.e. pass
to consider coordinates \beql{eq:rcc} y^i \ = \ \Phi^i (x,t;w) \ ,
\eeq leaving the time coordinate $t$ and the Wiener processes $w^k
(t)$ unaffected, the equation \eqref{eq:Ito} is mapped into a new
equation \beql{eq:Yito} d y^i \ = \ F^i \, d t \ + \ S^i_{\ k} \,
d w^k \ , \eeq where the new coefficients $F$ and $S$ are given by
\begin{eqnarray}
F^i &=& \frac{\pa \Phi^i}{\pa t} \ + \ f^j \, \frac{\pa \Phi^i}{\pa x^j}
 \ + \ \frac12 \, \Delta (\Phi^i) \ ,  \\
S^i_{\ k} &=& \frac{\pa \Phi^i}{\pa w^k} \ + \
 \s^j_{\ k} \,
\frac{\pa \Phi^i}{\pa x^j}  \ ; \end{eqnarray} see \cite{GL2,GS17}
for details of the computation.

It should be stressed that albeit the $F$ and $S$ are given here
as functions of the old variables $x$, as $f,\s$ and $\Phi^i$ all
depend of them, they should be thought as functions of the new
coordinates $y^i$ through the change of variables inverse to
\eqref{eq:rcc}, which we write as \beq x^i \ = \ \Theta^i (y,t;w)
\ . \eeq

The point is that in general the $F,S$ can and will depend not
only on the $(y,t)$ variables, but on the Wiener processes as well
(both through the explicit $w$-dependence of the $\Phi^i$ and
through the dependence of the $\Theta^i$ on the $w^k$). If this
happens, the new equation \eqref{eq:Yito} will \emph{not} be of
Ito type (see however Remark {\RKK}  in this respect).

Thus we will have a transformed equation \eqref{eq:Yito} again of
Ito type if and only if the additional conditions \beq \^\pa_m F^i
\ = \ 0 \ = \ \^\pa_m  S^i_{\ k} \eeq are satisfied for all
choices of $i$ and $k$ and for all $m$. In view of the explicit
expressions for $F$ and $S$ (see above), these conditions are also
written as
\begin{eqnarray}
& & \( \pa_t \ + \ f^j \, \pa_j \ + \ \frac12 \, \Delta \) \
\( \^\pa_m \, \Phi^i \) \ = \ 0 \ , \label{eq:I2I1} \\
& & \( \^\pa_k \ + \ \s^j_{\ k} \, \pa_j \) \ \( \^\pa_m \, \Phi^i
\) \ = \ 0 \ \label{eq:I2I2}  \end{eqnarray} (these represent a
generalization of the similar condition for the case of a scalar
equation, determined in \cite{GL2}).

In other words, we should require that all the components of the
gradient of $\Phi^i$ w.r.t. the Wiener coordinates $w^m$ belong to
the intersection of the kernels of the linear differential
operators\footnote{In \cite{GS17} we have proposed the name
``Misawa vector fields'', see there for the motivation of such a
name, for $Y_k = L_k$ and $Y_0 = L_0 - (1/2) \Delta$.}
\beql{eq:C6} L_0 \ := \ \pa_t \ + \ f^j \, \pa_j \ + \
\frac12 \, \Delta \ \ \ ; \ \ \ L_k \ := \ \^\pa_k \ + \ \s^j_{\
k} \, \pa_j \ . \eeq

Needless to say, \eqref{eq:C6} is always satisfied when $\Phi$
does not depend on the $w^k$ variables, i.e. for deterministic
changes of variables.

We can then easily extend Proposition 4 to the following one, in
which we make free use of the notation established in Remark \LDS.

\medskip\noindent
{\bf Proposition 5.} {\it Suppose the system \eqref{eq:Ito} admits
an $r$-parameter \emph{solvable} Lie algebra $\mathcal{G}$ of
simple -- deterministic or random -- symmetries, with generators
\beq \mathbf{X}_{(k)} =\sum_{i=1}^n \varphi_k^i(x,t) \
\frac{\pa}{\pa x_i} \qquad (k=1,...,r) \ , \eeq acting regularly
with $r$-dimensional orbits. Let these be labeled according to the
derived series for $\mathcal{G}$, and let $E_q$ be the equation
obtained after reduction by the first $q$ symmetries.

Suppose moreover that, with $\Phi^i_{(k)}$ the maps \eqref{eq:rcc}
describing the change of variables associated to the symmetries
$X_{(k)}$, the equations \eqref{eq:I2I1}, \eqref{eq:I2I2} are
satisfied for equation $E_{k-1}$.

Then the system \eqref{eq:Ito} can be reduced to a system of $m =
(n-r)$ Ito equations, \beq d y^i \ = \ g^i (y^1,...,y^m;t) \, d t \ +
\ \s^i_{\ k} (y^1,...,y^m;t) \, d w^k \ \ \ \ (i,k=1,...,m) \eeq
and $r$ ``reconstruction equations'', the solutions of which can
be obtained by (stochastic) quadratures from the solution of the
reduced $(n-r)$-order system.}

\bigskip\noindent
{\bf Proof.} If equations \eqref{eq:I2I1} and \eqref{eq:I2I2} are
satisfied, we are guaranteed Ito equations are mapped into Ito
equations, thus the application of each map associated to
symmetries $X_{(k)}$ transform the equation $E_{k-1}$ (and in
particular the original system \eqref{eq:Ito}) into one of the
same nature (and dimension). Moreover Proposition 1 guarantees
that after the application of the map, $X_{(k)}$ is still a
symmetry of the new system.

Thus the new system is still of Ito type but with r.h.s. not
depending on one of the $x^i$ variables, and if the maps are performed
in the proper order, i.e. following the Lie algebraic structure of
the symmetry algebra, at each step we eliminate an additional
variable with no risk of reintroducing dependencies on previously
eliminated ones.

Alternatively, once we are guaranteed to remain within the class
of Ito equations, we can deal with the associated Stratonovich
ones, which admit the same symmetries \cite{GL1,Unal} and
transform according to the standard chain rule. We can then
proceed as in the case of deterministic equations, and reach the
same conclusion, modulo the substitution of standard integrals by
stochastic ones in the reconstruction equations. \EOP
\bigskip

\medskip\noindent
{\bf Remark 12.} Note that the conditions \eqref{eq:I2I1} and
\eqref{eq:I2I2} should be checked at each step of the reduction
procedure. We do not have determined a criterion to establish
\emph{apriori} -- i.e. just on the original system -- if this will
be the case, at least to some order. We also remark that albeit we
have seen that reduction can be effective even if it leads us
outside the realm of Ito equations (see Remarks {\RK} and {\RKK}),
if this is the case for intermediate equations we are not
guaranteed the symmetries will be preserved in the reduction
procedure. In fact, this results rests on the relation between
symmetries of Ito and the equivalent Stratonovich equation, and
the matter has not been investigated (neither here nor elsewhere
in the literature) for non-Ito equations. \EOR


\section{W-maps and W-symmetries}
\label{sec:Wsymm}

The definition of simple symmetries can be too restrictive even
for obviously invariant systems. We will consider one of these,
i.e. the isotropic (linear or non linear) stochastic oscillators,
as a motivating example.

\subsection{Stochastic oscillators}
\label{sec:motivate}

Consider e.g. the isotropic ``stochastic harmonic oscillator''
($i=1,...,n$): \beql{eq:WROTlin} d x^i \ = \ - x^i \, dt \ + \ d
w^i \ , \eeq or more generally the system \beql{eq:WROT} d x^i \ =
\ - F(|\xb|^2) \, x^i \, dt \ + \ S (|\xb|^2) d w^i \ , \eeq where
$F$ and $S$ are scalar functions of $|\xb|^2 = (x^1)^2 + ...
+(x^n)^2$ alone, and there are as many independent Wiener
processes as $x$ variables. In view of our definition, this is
\emph{not} rotationally invariant (under standard, deterministic
or random, maps): this is due to the fact that we can rotate the
$\xb = (x^i,...,x^n)$ vector, but we are \emph{not} allowed to
rotate at the same time also the ${\bf w} = (w^i,...,w^n)$ one.

On the other hand, \emph{if} we consider maps also acting on the
$w$ variables, then \eqref{eq:WROT} \emph{is} obviously invariant
under simultaneous (identical) rotations in the $x$ \emph{and} in
the $w$ variables spaces.

Similarly, \eqref{eq:WROTlin} -- but not \eqref{eq:WROT}, in
general -- is also invariant under a simultaneous identical
scaling of the $x^i$ and the $w^i$ variables, $x^i \to \la x^i$,
$w^i \to \la w^i$.
(This Example will be considered in more detail below.)

\subsection{W maps}

It is thus natural to consider also maps acting on the Wiener
processes themselves, $w \to z$. At first sight, as we want to have
again standard independent Wiener processes (we want to have, in
the end, an equation of the same type as the original one, let
alone this being exactly the same), the $z^i$ can be at most of
the form \beq z^i \ = \ R^i_{\ j} \ w^j \ , \eeq with $z^m$
independent unit Wiener processes and $R$ a \emph{constant
orthogonal} matrix, $R \in \mathrm{O} (n)$. (This point of
view was used in \cite{GS17}; the discussion given there should be
amended with the considerations to follow.)

But this is not entirely correct. In fact, as for rescalings of
time, we can allow to obtain a non-standard Wiener process
provided the non-standard nature amounts to a scalar (not
necessarily constant) factor, which can then be adsorbed by the
coefficients $S^i_k$.

On the other hand, it is essential to preserve the independence of
the Wiener processes; in geometrical terms this means that the
transformation must preserve the (right) angle between the
different Wiener processes.

In other words, we must consider \emph{conformal} transformations,
possibly depending on the $({\bf x},t)$ variables. As we will see,
these will be subject to further constraints.

\medskip\noindent
{\bf Remark 13.} It is well known that the conformal group in a
$d$-dimensional space (with $d \not= 2$; here we will not dwell
into the special properties of the conformal group for $d=2$) is
made of translations, certain linear transformations, more
precisely orthogonal ones and dilations, and certain quadratic
transformations, also known as \emph{special conformal maps}
(which are singular in the origin). In our context, obviously
translations should be discarded (they would produce stochastic
processes which have non-zero average increment and hence are not
even a martingale); and we are not willing to admit singular maps
or to forbid the process to go through zero. So we are left with
linear conformal maps alone.

It is also known that the conformal group in $d \not= 2$
dimensions is isomorphic to the group $SO(d+1,1)$, which gives a
practical way to tackle the case of general conformal group,
albeit the isomorphism can introduce some computational
difficulties.
\EOR
\bigskip

In the present work, for the reasons sketched in Remark 13, we
will \emph{not} consider translations nor special conformal maps,
and restrict our attention to the simplest sector of linear
conformal maps, i.e. rotations and dilations. It will turn out
that these have different standings in the present context, see
Section \ref{sec:RWSIS} below.

We will thus consider in general transformations of the type
$$
x^i = \Phi^i (y,\theta ;z) \ , \ \ t = \Theta (\theta) \ , \ \ w^k
= R^k_{\ m} (y,\theta ) \, z^m \ , $$ with $R = R (y,\theta) \in
O(n) \times {\bf R}_+$, which we will denote as the linear
conformal group. Moreover, we should require $\Theta' (t) > 0$ for
all $t$.

Actually, we know that in Kozlov theory only vector fields with no
component along $t$ are of interest, so from now on we will assume
$\Theta (t) \equiv t$, and the considered transformations will
just be (with again $R \in O(n) \times {\bf R}_+$)
\begin{eqnarray}
x^i &=& \Phi^i (y,t ;z) \ , \nonumber \\
w^k &=& h(y,t;z) \ = \ R^k_{\ m} (y,t ) \, z^m \ .
\label{eq:WmapK0}
\end{eqnarray}
We will refer to these as \emph{linear
W-maps}, and correspondingly we may have \emph{linear
W-symmetries}. Note that ``linear'' only refers to the action on
the sector of the $w$ variables.

The inverse of this map will be written as
\begin{eqnarray}
y^i &=& \Psi^i (x,t ;w) \ , \nonumber \\
z^k &=& A^k_{\ m} (x,t ) \, w^m \ ; \label{eq:WmapK0inv}
\end{eqnarray} here $A$ is again in the linear conformal group. As
mentioned above, the new Wiener processes should be independent,
i.e. we should require
$$ d z^i \cdot d z^j \ = \ \delta^{ij} \ \zeta (x,t) \, d t \ ; $$
it is essential that $\zeta$ should not depend on the $w$, albeit
it is -- in principles, but see below -- allowed to depend on the
$x,t$ variables.

By a straightforward application of Ito rule (and, in the last
step, restricting to the solutions to \eqref{eq:Ito}) and up to
terms of order $o (dt)$, we get
\begin{eqnarray*}
d z^p &=& (\pa_k A^p_{\ j}) \, w^j \ d x^k \ + \ (\pa_t A^p_{\ j}) \, w^j \ d t
\ + \ A^p_{\ j} \ d w^j \ + \ \frac12 \Delta (A^p_{\ j} w^j ) \ d t \ ; \\
d z^p \cdot d z^q &=& [A^p_{\ j} A^q_{\ k}] \ (d w^j \cdot d w^k)
\ + \ [(\pa_i A^p_{\ j} ) w^j (\pa_k A^q_{\ \ell} ) w^\ell ] \ (d x^i \cdot d x^k ) \\
& & \ + \ [(\pa_i A^p_{\ j}) w^j \, A^q_{\ k}] \ (d x^i \cdot d
w^k) \ + \
[A^p_{\ i} \, (\pa_j A^q_{\ \ell} \, w^\ell] \ (d w^i \cdot d x^j)  \\
&=& [A^p_{\ j} A^q_{\ k}] \ \delta^{jk} \ d t  \ + \
[(\pa_i A^p_{\ j} ) w^j (\pa_k A^q_{\ \ell} ) w^\ell ] \
(\s^i_{\ r} \s^k_{\ s} \delta^{rs} ) \ d t  \\
& & \ + \ [(\pa_i A^p_{\ j}) w^j \, A^q_{\ k}] \ (\s^i_{\ r}
\delta^{kr}) \ d t \ + \ [A^p_{\ k} \, (\pa_j A^q_{\ \ell} \,
w^\ell] \ (\sigma^j_{\ s} \delta^{k s}) \ d t  \ .
\end{eqnarray*}

Thus, in order to have $\zeta = \zeta (x,t)$ as required, we must
impose that {\it the $A$ matrices do not depend on the spatial
variables $x$}, i.e. $A = A(t)$; and hence, recalling that $R
=A^{-1}$, also that {\it the $R$ do not depend on the new spatial
variables $y$}, i.e. $R = R(t)$.

We are thus reduced to consider maps \eqref{eq:WmapK0} of the form
\begin{eqnarray}
x^i &=& \Phi^i (y,t ;z) \ , \nonumber \\
w^k &=& h^m (t,z) \ = \ R^k_{\ m} (t) \, z^m \ , \label{eq:WmapK1}
\end{eqnarray} with $R (t)$ in the linear conformal group.

Note that with this form of $h^m$, we immediately have
\beql{eq:Delta_h} \Delta (h^m ) \ = \ 0 \ . \eeq Moreover,
expressing $d w^k$ in terms of the new variables we get
\beql{eq:dwkWW0} d w^k \ = \ R^k_m \, d z^m \ + \ (\pa_t R^k_m) \,
z^m \, d t \ . \eeq

This is not acceptable: in fact we know that a Wiener process has
$d w \simeq \sqrt{d t}$. Thus the last term in \eqref{eq:dwkWW0}
must be zero, i.e. $\pa_t R^k_m = 0$, i.e. $R$ can not depend on
$t$.

Thus in conclusion, summarizing our discussion in a  formal
statement, we have:

\medskip\noindent
{\bf Lemma 1.} {\it Acceptable W-maps \eqref{eq:WmapK0} have
\beql{eq:HRW} h^m = R^m_{\ k} w^k\ , \eeq and therefore
\beql{eq:dwkWW} d w^k \ = \ R^k_m \, d z^m \ . \eeq With this,
\eqref{eq:WmapK1} further reduce to
\begin{eqnarray}
x^i &=& \Phi^i (y,t ;z) \ , \nonumber \\
w^k &=& A^k_{\ m} \, z^m \ , \label{eq:WmapK}
\end{eqnarray} with $A$ a constant matrix in the linear conformal
group.}
\bigskip

When we consider infinitesimal maps, we have
\begin{eqnarray*}
x^i &=& \Phi^i (y,t ;z) \ = \ y^i \ + \ \eps \, \vphi^i (y,t;z) \ , \\
w^k &=& A^k_{\ m} \, z^m \ = \ z^k \ + \ \eps \, R^k_{\ m} \, z^m
\ , \end{eqnarray*} hence these will be generated by vector fields
\beql{eq:WmapKX} X \ = \ \vphi^i (x,t;w) \, \pa_i \ + \ \( R^k_{\
m} \, w^m  \) \, \^\pa_k \ . \eeq

\subsection{W symmetries}

We will then proceed as usual in order to determine the effect of
the map \eqref{eq:WmapK} on the Ito equation \eqref{eq:Ito}. It
follows from \eqref{eq:WmapK} that (we stress that $\pa_i$,
$\^\pa_k$ and $\Delta$ are now defined w.r.t. the new set of
variables) \beq d x^i \ = \ (\pa_t \Phi^i) \, d t \ + \ (\pa_j
\Phi^i) \, d y^j \ + \ \frac12 (\Delta \Phi^i) \, d t \ + \
(\^\pa_m \Phi^i) \, d z^m \ . \eeq Comparing this with the Ito
equation under study \eqref{eq:Ito} and writing for ease of
notation \beql{eq:M00} M^i_{\ j} \ := \ \frac{\pa \Phi^i}{\pa y^j} \ ,
\eeq we readily
obtain \beql{eq:We1} M^i_{\ j} \, d y^j \ = \ \( \wt{f}^i - \pa_t
\Phi^i - \frac12 \Delta \Phi^i \) \, d t \ + \ \wt{\s}^i_{\ k} \,
d w^k \ - \ ( \^\pa_m \Phi^i ) \, d z^m \ . \eeq We stress that
now $f$ and $\s$ should be thought as functions of the new
variables; thus we introduced
$$ \wt{f}^i (y,t;z) \ := \ f^i [\Phi(y,t;z),t] \ , \ \
\wt{\s}^i_{\ k} (y,t:z) \ := \ \s^i_{\ k} [\Phi(y,t;z),t] \ . $$

Note that by assumption $M$ is invertible, as the map
\eqref{eq:WmapK} provides a change of variables; we will denote
the inverse of $M$ by $\La$, \beql{eq:La} \La^i_{\ j} \ = \
\frac{\pa y^i}{\pa x^j} \ . \eeq

In \eqref{eq:We1} we should still express $d w^k$ in the new
variables, i.e. use  \eqref{eq:dwkWW}. Inserting this into
\eqref{eq:We1} we get \beql{eq:We2} M^i_{\ j} \ d y^j \ = \
\(\wt{f}^i - \pa_t \Phi^i - \frac12 \Delta \Phi^i \)  \ d t \ + \
\[ (\^\pa_m \Phi^i) \ + \  \wt{\s}^i_{\ k} \, R^k_{\ m} \] \  d
z^m \ . \eeq Multiplying by $M^{-1}=\La$, \eqref{eq:We2} yields
finally
\begin{eqnarray}
d y^i &=& \La^i_j \, \[ \wt{f}^j - \pa_t \Phi^j - \frac12 \Delta
\Phi^i \] \, d t
\ + \  \La^i_j \, \[ \^\pa_m \Phi^j + \wt{\s}^j_k R^k_m \] \, d z^m \nonumber \\
&:=& F^i \, d t \ + \ S^i_{\ m} \, d z^m \ , \label{eq:We3}
\end{eqnarray} where we have of course introduced the compact
notation \beql{eq:WFS} F^i \ = \ \La^i_j \, \( \wt{f}^j - \pa_t
\Phi^j - \frac12 \Delta \Phi^j \)  \ , \ \ S^i_{\ m} \ = \ \La^i_j
\, \[ \^\pa_m \Phi^j + \wt{\s}^j_k R^k_m \] \ . \eeq

\medskip\noindent
{\bf Remark 14.} For \eqref{eq:We3} to be again an Ito equation,
we need that both the conditions \beql{eq:GRCV} (\pa F^i / \pa
z^\ell) = 0 \ , \ \ \ (\pa S^i_m / \pa z^\ell) = 0 \eeq hold, for
all $i,m,\ell$. These equations provide the further limitation on
the form of $\Phi$ and $h$, i.e. $R$, mentioned above. On the
other hand, as already remarked (see in particular Remarks {\RK}
and {\RKK}), the requirement to stay within the class of Ito
equations can be too restrictive for a number of concrete
applications. \EOR

\subsection{Split W-symmetries}
\label{sec:Wsplit}


We note that in the case we have a change of variables of the
simpler form
\begin{eqnarray}
x^i &=& \Phi^i (y,t) \ , \nonumber \\
w^k &=& R^k_{\ m} \, z^m \ , \label{eq:WmapSK1} \end{eqnarray}
i.e. when the change of variables does not mix the spatial
variables and the Wiener processes, the situation is substantially
simpler.

In fact, now $M$, $\La$, $\wt{f}$ and $\wt{\s}$ are all
independent of $z$, and both equations \eqref{eq:GRCV} are always
satisfied.

We have thus identified a simple class of $W$-maps,
\eqref{eq:WmapSK1}, which is guaranteed to map Ito equations into
Ito equations.

As in this case the spatial variables and the Wiener process
transform independently of each other, we will refer to this class
of maps as \emph{split W-maps}; our discussion above shows that:

\medskip\noindent
{\bf Lemma 2.} {\it Split W-maps transform Ito equations into Ito
equations.}
\bigskip

If some split W-maps leave a given equation invariant, we will
speak of \emph{split W-symmetries}.

\medskip\noindent
{\bf Remark 15.} We anticipate and stress that the (Kozlov-type)
change of variables associated to a split W-symmetry rectifying it
-- see Section \ref{sec:Wkoz} -- is in general \emph{not} a split
W-map. \EOR

\medskip\noindent
{\bf Remark 16.} It is interesting to note that vector fields,
including symmetries, transform in a specially simple way under
split W-maps. In fact, in general
\begin{eqnarray*} \frac{\pa}{\pa x^i} &=& \frac{\pa y^j}{\pa x^i} \, \frac{\pa }{\pa y^j}
\ + \ \frac{\pa z^m}{\pa x^i} \, \frac{\pa }{\pa z^m} \ , \\
\frac{\pa}{\pa w^k} &=& \frac{\pa y^j}{\pa w^k} \, \frac{\pa }{\pa
y^j} \ + \ \frac{\pa z^m}{\pa w^k} \, \frac{\pa }{\pa z^m} \ ;
\end{eqnarray*} actually, as we have seen that in admissible maps
$z$'s do not depend on $x$'s, \beql{eq:WKvfgen0} \frac{\pa}{\pa
x^i} \ = \ \frac{\pa y^j}{\pa x^i} \, \frac{\pa }{\pa y^j} \ , \ \
\frac{\pa}{\pa w^k} \ = \ \frac{\pa y^j}{\pa w^k} \, \frac{\pa
}{\pa y^j} \ + \ \frac{\pa z^m}{\pa w^k} \, \frac{\pa }{\pa z^m} \
. \eeq

Moreover, for split W-maps we also have $\pa y / \pa w = 0$, and
in general $\pa y / \pa x = \La$; hence these reduce to
\beql{eq:WKvfsplit0} \frac{\pa}{\pa x^i} \ = \ \La^j_{\ i} \,
\frac{\pa }{\pa y^j} \ , \ \ \frac{\pa}{\pa w^k} \ = \ \frac{\pa
z^m}{\pa w^k} \, \frac{\pa }{\pa z^m} \ . \eeq Note also that,
with $Q = R^{-1}$, it follows from \eqref{eq:WmapK} -- hence for
all kind of W-maps -- that $ \pa z^m / \pa w^k = Q^m_{\ k}$; thus
\eqref{eq:WKvfgen0} and \eqref{eq:WKvfsplit0} are respectively
\beql{eq:WKvfgen} \frac{\pa}{\pa x^i} \ = \ \La^j_{\ i} \,
\frac{\pa }{\pa y^j} \ , \ \ \frac{\pa}{\pa w^k} \ = \ \frac{\pa
y^j}{\pa w^k} \, \frac{\pa }{\pa y^j} \ + \ Q^m_{\ k} \, \frac{\pa
}{\pa z^m} \eeq in the general case; and \beq \frac{\pa}{\pa x^i}
\ = \ \La^j_{\ i}  \, \frac{\pa }{\pa y^j} \ , \ \ \frac{\pa}{\pa
w^k} \ = \ Q^m_{\ k} \, \frac{\pa }{\pa z^m} \ . \eeq in the split
one. \EOR

\medskip\noindent
{\bf Remark 17.} Note that the simultaneous rotations in $x$ and
in $w$ space considered in Sect.\ref{sec:motivate} correspond to a
split W-symmetry.  \EOR

\section{W-symmetries of Ito versus associated Stratonovich equations}
\label{sec:WSIS}

The main tool allowing for an effective description and use of
standard (deterministic or random) admissible symmetries of Ito
equation is the result identifying these with symmetries of the
associated Stratonovich equation; see Section \ref{sec:geom}
above.

We thus wonder if a similar result also holds for $W$-symmetries;
or at least for split W-symmetries, or under some additional
condition. The present Section provides an answer to this
question.

\subsection{Determining equations for W-symmetries of Ito equations}

We consider the Ito equation \eqref{eq:Ito} and act on it by the
simple vector field \beql{eq:X} X \ = \ \vphi^i (x,t;w) \, \pa_i \
+ \ h^k (x,t;w) \, \^\pa_k  \eeq (note that at the moment we are
not restricting the form of $h^k$; see Remark 19 below about
this).

The action of $X$ is described by
\beql{eq:Xmap} x^i \ \to \ x^i \ + \ \eps \, \vphi^i (x,t;w) \ , \ \
w^k \ \to \ w^k \ + \ \eps \, h^k (x,t;w) \ , \eeq while $t$ remains unaffected.

With standard computations, using also \eqref{eq:Ito} itself, we
obtain that, at first order in $\eps$,
\begin{eqnarray*}
d x^i & \to & d x^i \ + \ \eps \ \[ (\pa_t \vphi^i) dt + (\pa_j \vphi^i) d x^j
+ (\^\pa_k \vphi^i) d w^k + \frac12 \Delta (\vphi^i) d t \] \\
&=& d x^i \ + \ \eps \ \[ \( \pa_t \vphi^i + f^j \pa_j \vphi^i +
\frac12 \Delta \vphi^i \) \, d t \ + \
\( \^\pa_k \vphi^i + \s^j_{\ k} \pa_j \vphi^i \) \, d w^k \] \ , \\
d w^k & \to & d w^k \ + \ \eps \ \[ (\pa_t h^k) d t + (\pa_j h^k ) d x^j
+ (\^\pa_m h^k) d w^m + \frac12 (\Delta h^k) d t \] \\
&=& d w^k \ + \ \eps \ \[ \( \pa_t h^k + f^j \pa_j h^k + \frac12 \Delta h^k \) \, d t
\ + \ \( \^\pa_m h^k + \s^j_{\ m} \pa_j h^k \) \, d w^m \] \ ; \\
f^i & \to & f^i  \ + \ \eps \ \vphi^j \, \pa_j f^i  \ , \\
\s^i_{\ k} & \to & \s^i_{\ k} \ + \ \eps \ \vphi^j \, \pa_j \s^i_{\ k} \ . \end{eqnarray*}

With these and some standard computations, it is easy to check
that the condition for the equation to remain invariant is that
the following equations hold for all $i$ and $k$:
\begin{eqnarray}
\pa_t \vphi^i \, + \, (f^j \pa_j \vphi^i  - \vphi^j \pa_j f^i) \,
+ \, \frac12 \, \Delta \vphi^i &=&
\s^i_{\ k} \, \( \pa_t h^k +  f^j \pa_j h^k  +  \frac12 \Delta h^k \) \ , \label{eq:WIde10} \\
\^\pa_k \vphi^i \, + \, (\s^j_{\ k} \pa_j \vphi^i  -  \vphi^j
\pa_j \s^i_{\ k} )  &=& \s^i_{\ m} \, \( \^\pa_k h^m  +  \s^j_{\
k} \pa_j h^m \) \ . \label{eq:WIde20} \end{eqnarray}

These were obtained for a \emph{general} $h$; but we have seen in
Lemma 1 that $h$ is of the form \eqref{eq:HRW}, hence the above
equations further reduce and we have:

\medskip\noindent
{\bf Lemma 3} {\it The \emph{determining equations} for (general
simple) $W$-symmetries of the Ito equation \eqref{eq:Ito} are
\begin{eqnarray}
\pa_t \vphi^i \ + \ (f^j \pa_j \vphi^i \, - \, \vphi^j \pa_j f^i)
\ + \ \frac12 \, \Delta \vphi^i & = & 0 \ ,
\label{eq:WIde1} \\
\^\pa_k \vphi^i \ + \ (\s^j_{\ k} \pa_j \vphi^i \, - \, \vphi^j
\pa_j \s^i_{\ k} ) \ - \ \s^i_{\ m} \ R^m_{\ k} & = & 0 \ .
\label{eq:WIde2}
\end{eqnarray} }

\subsection{Determining equations for W-symmetries of Stratonovich equations}

The computations are pretty much similar -- at the exception of
using the chain rule rather than the Ito one -- when considering
the Stratonovich equation \beql{eq:Strat} d x^i \ = \ b^i (x,t) \,
d t \ + \ \s^i_{\ k} (x,t) \circ d w^k \ ; \eeq in this case we
obtain the determining equations for (general simple)
$W$-symmetries of the Stratonovich equation \eqref{eq:Strat} in
the form
\begin{eqnarray}
\pa_t \vphi^i &+& (b^j \pa_j \vphi^i \, - \, \vphi^j \pa_j b^i) \ = \
\s^i_{\ k} \ \( \pa_t h^k \, + \, b^j \pa_j h^k \) \ , \label{eq:WSde10} \\
\^\pa_k \vphi^i &+& (\s^j_{\ k} \pa_j \vphi^i \, - \, \vphi^j
\pa_j \s^i_{\ k} ) \ = \ \s^i_{\ m} \ \( \^\pa_k h^m \, + \,
\s^j_{\ k} \pa_j h^m \) \ . \label{eq:WSde20} \end{eqnarray}.

We note immediately that \eqref{eq:WSde20} coincides with
\eqref{eq:WIde2}, and this with \emph{general} form of $h^k$. For
$h$ as dictated by Lemma 1, i.e. as in \eqref{eq:HRW}, the above
equations are simplified, and we get:

\medskip\noindent
{\bf Lemma 4.} {\it The \emph{determining equations} for (general
simple) $W$-symmetries of the Stratonovich equation
\eqref{eq:Strat} are
\begin{eqnarray}
\pa_t \vphi^i &+& (b^j \pa_j \vphi^i \, - \, \vphi^j \pa_j b^i) \ = \ 0 \ , \label{eq:WSde1} \\
\^\pa_k \vphi^i &+& (\s^j_{\ k} \pa_j \vphi^i \, - \, \vphi^j
\pa_j \s^i_{\ k} ) \ = \ \s^i_{\ m} \ R^m_{\ k} \ .
\label{eq:WSde2} \end{eqnarray}. }

\subsection{The relation between symmetries of an Ito and of the
associated Stratonovich equations}
\label{sec:RWSIS}

In order to compare symmetries of the equations \eqref{eq:Ito} and
\eqref{eq:Strat}, we should require that \eqref{eq:Strat} is just
the equation associated to \eqref{eq:Ito}. As well known, this
amounts to requiring that \beql{eq:rhosigma} f^i (x,t) \ = \ b^i
(x,t) \ + \ \rho^i (x,t) \ ; \ \ \ \rho^i \ = \ \frac12 \, (\pa_k
\s^{ij} ) \, \s^k_{\ j} \ . \eeq

Using this, \eqref{eq:WIde10} is rewritten as \begin{eqnarray*}
\pa_t \vphi^i &+& (b^j \pa_j \vphi^i - \vphi^j \pa_j b^i) \ + \
(\rho^j \pa_j \vphi^i -\vphi^j \pa_j \rho^i ) \ + \ \frac12 \Delta
\vphi^i \\ &=& \s^i_{\ k} \, \( \pa_t h^k + b^j \pa_j h^k + \rho^j
\pa_j h^k + \frac12 \Delta h^k \) \ . \end{eqnarray*} Subtracting
\eqref{eq:WSde10} from this, we get \beq (\rho^j \pa_j \vphi^i -
\vphi^j \pa_j \rho^i ) \ + \ \frac12 \Delta \vphi^i \ = \ \s^i_{\
k} \, \( \rho^j \pa_j h^k + \frac12 \Delta h^k \) \ ; \eeq
recalling the definition of $\rho$ in terms of $\s$, see
\eqref{eq:rhosigma}, this reads \beq (\pa_k \s^{jm} ) \, \s^k_{\
m} \, (\pa_j \vphi^i) \ - \ \vphi^j \pa_j [(\pa_k \s^{im} ) \,
\s^k_{\ m}] \ + \ \Delta \vphi^i \ = \ \s^i_{\ k} \[ (\pa_q
\s^{jm} ) \, \s^q_{\ m} \, (\pa_j h^k) + \Delta h^k \] \ . \eeq

Moreover, if we restrict to linear W-symmetries, $h = R z$, the
terms in the square bracket on the r.h.s.\footnote{These are the
only ones depending on $h$, so apparently we obtain a condition
which is independent of $h$ when this satisfies the condition set
in Lemma 1; but a dependence on $h$ will be introduced when we
restrict to solutions to the common set of determining equations
\eqref{eq:WIde2}, \eqref{eq:WSde2}, see below.} both vanish, and
we are just left with \beql{eq:WS1} (\pa_k \s^{jm} ) \, \s^k_{\ m}
\, (\pa_j \vphi^i) \ - \ \vphi^j \pa_j [(\pa_k \s^{im} ) \,
\s^k_{\ m}] \ + \ \Delta \vphi^i \ = \ 0 \ , \eeq which we also
write as \beql{eq:WS1B} \Delta (\vphi^i ) \ = \ \Sigma ( \vphi^i )
\ , \eeq having of course defined \beql{eq:Sigma} \Sigma ( \vphi^i
) \ := \ \vphi^j \pa_j [(\pa_k \s^{im} ) \, \s^k_{\ m}] \ - \
(\pa_k \s^{jm} ) \, \s^k_{\ m} \, (\pa_j \vphi^i) \ . \eeq

We stress that in order to have the same symmetries for the Ito
and the associated Stratonovich equations, it is not required that
\eqref{eq:WS1} holds in general, but only that it holds when
\eqref{eq:WSde2} is also satisfied. That is, we can substitute in
\eqref{eq:WS1} -- in particular, in the term $\Delta \vphi^i$ --
for the derivatives of $\vphi^i$ (involving derivation w.r.t. at
least one Wiener process) according to \eqref{eq:WSde2} and its
differential consequences. Recalling that $\s$ does not depend on
the $w^k$ and that $R$ is constant, we get:
\begin{eqnarray*}
\^\pa_k \vphi^i &=& (\vphi^j \pa_j \s^i_{\ k} \ - \ \s^j_{\ k} \pa_j \vphi^i )
\ + \ \s^i_{\ m} \ R^m_{\ k} \ , \\
\pa_\ell \^\pa_k \vphi^i &=& \( (\pa_\ell \vphi^j) (\pa_j \s^i_{\ k})
\ + \ \vphi^j (\pa_\ell \pa_j \s^i_{\ k} ) \ - \ (\pa_\ell \s^j_{\ k}) (\pa_j \vphi^i )
\ - \ \s^j_{\ k} (\pa_\ell \pa_j \vphi^i) \) \\ & & \ + \ (\pa_\ell \s^i_{\ m} ) \ R^m_{\ k} \ ; \\
\^\pa_m \^\pa_k \vphi^i &=& \( (\^\pa_m \vphi^j) (\pa_j \s^i_{\ k})
\ - \ \s^\ell_{\ k} (\^\pa_m \pa_\ell \vphi^i) \)  \\
&=& [(\vphi^p \pa_p \s^j_{\ m} \ - \ \s^p_{\ m} \pa_p \vphi^j )
\ + \ \s^j_{\ q} \ R^q_{\ m} ] (\pa_j \s^i_{\ k} ) \\ & & \ - \ \s^j_{\ k} \,
\[ (\pa_j \vphi^p) (\pa_p \s^i_{\ m} ) \ + \ \vphi^p (\pa_j \pa_p \s^i_{\ m} )
\ - \ (\pa_j \s^p_{\ m} ) (\pa_p \vphi^i ) \right. \\
& & \left. \ \ \ - \ \s^p_{\ m} (\pa_j \pa_p \vphi^i )
\ + \ (\pa_j \s^i_{\ q} ) \ R^q_{\ m} \]  \ . \end{eqnarray*}

We can now insert these expressions into the explicit form
\eqref{eq:ItoLapl} of $\Delta \vphi$. \footnote{Note that here we
will raise and lower indices -- also inside spatial derivatives --
making use of the assumption we are working in an Euclidean space;
it would be interesting to study if one obtains different results
in a general Riemannian manifold.}. One obtains in this way
\begin{eqnarray*}
\De \vphi^i &=& \s^{jk} \s^m_{\ k} \pa_j\pa_m \vphi^i \ + \ 2
\s^{\ell k} \[ (\pa_\ell \vphi^j)(\pa_j \s^i_{\ k}) + \vphi^j
(\pa_\ell \pa_j \s^i_{\ k}) \right. \\
& & \ \ \left. - (\pa_\ell \s^j_{\ k} )(\pa_j \vphi^i) - \s^j_{\
k} (\pa_\ell \pa_j \vphi^i) + (\pa_\ell \s^i_{\
p}) R^p_{\ k} \] \\
& & + \ \delta^{\ell k} \[ \vphi^p (\pa_p \s^j_{\ \ell} ) -
\s^p_{\ \ell} (\pa_p \vphi^j) + \s^j_{\ q} R^q_{\ \ell} \] (\pa_j
\s^i_{\ k}) \\
& & - \ \delta^{\ell k} \, \s^j_{\ k} \[ (\pa_j \vphi^p) (\pa_p
\s^i_{\ \ell} ) + \vphi^p (\pa_j \pa_p \s^i_{\ \ell} ) - (\pa_j
\s^p_{\ \ell}) (\pa_p \vphi^i) \right. \\
& & \ \ \left. - \s^p_{\ \ell} (\pa_j \pa_p
\vphi^i) + (\pa_j \s^i_{\ q} ) R^q_{\ \ell} \] \\
&=& \[ \s^{\ell k} (\pa_\ell \pa_j \s^i_{\ k}) + (\pa_j \s^{\ell
k} )(\pa_\ell \s^i_{\ k})\] \vphi^j \ - \ \s^{jk} (\pa_j \s^p_{\
k} ) (\pa_p \vphi^i) \\
& & + \ \[ \s^{\ell k} (\pa_\ell \s^i_{\ p}) - \s^\ell_{\ p}
(\pa_\ell \s^{ik}) \] \, R^p_{\ k} \\
&=& \vphi^j \, \pa_j [ \s^{\ell k} (\pa_\ell \s^i_{\ k}) ] \ - \ [
\s^{\ell k} (\pa_\ell \s^j_{\ k})] \, \pa_j \vphi^i \ + \ \s^{\ell
k} (\pa_\ell \s^{ip} ) \, [R_{pk} - R_{kp} ] \\
&=& \Sigma (\vphi^i ) \ + \ \s^{\ell k} (\pa_\ell \s^{ip} ) \,
[R_{pk} + R_{kp} ] \ . \end{eqnarray*} In the last step we have
used the definition \eqref{eq:Sigma} of $\Sigma (\vphi^i)$.

We denote by \beql{eq:mathR} \mathcal{R} (\s) \ := \ \s^{\ell k}
(\pa_\ell \s^{ip} ) \, [R_{pk} + R_{kp} ] \eeq the term depending
on $R$ in the final result above; note that $\mathcal{R} (\s )$ is
identically zero for constant $\s$. In general \eqref{eq:WS1} is
satisfied if and only if $\mathcal{R} (\s) = 0$ on solutions to
\eqref{eq:WSde2}. (We recall that our computation was indeed based
on restricting to solutions to \eqref{eq:WSde2}.)

As already mentioned, the generators of the linear conformal
groups correspond to rotations and dilations; it is also well
known that the generators of rotations are skew-symmetric
matrices, while the generator of dilations is a diagonal
matrix.\footnote{We recall that we are considering infinitesimal
(near-identity) maps, see \eqref{eq:Xmap}; hence we are dealing
with the connected component of the identity in the group, and
with generators.}

We conclude immediately that if only the rotation part is present
in $R$, then $\mathcal{R} (\vphi^i) = 0$, while if the dilations
are also present we have in general, unless $\s$ satisfies
\beql{eq:Sdil} \[ \s^{jm} \, (\pa_j \s^{i}_{\ q} ) \ + \ \s^j_{\
q} \, (\pa_j \s^{i m} ) \] \, R^q_{\ m} \ = \ 0 \ , \eeq that
$\mathcal{R} (\vphi^i) \not= 0$ and hence we \emph{can} have a
difference between symmetries of the Ito and the associated
Stratonovich equations. This will be explicitly shown to be the
case in Example 5 below.

Note that \eqref{eq:Sdil} is always satisfied (for whatever $R$)
if and only if $\s$ is constant w.r.t. the spatial variables
$x^i$. Actually, as we know that the only ``dangerous'' situation
is that with $R$ diagonal (generating dilations), it is immediate
to check that in this case we have $\mathcal{R}=0$ if and only if
$\s$ is spatially constant.

We also recall that we have considered maps not acting on time;
thus it is not surprising that time derivatives of $\s$ do not
appear to play a role.

We can summarize our discussion as follows

\medskip\noindent
{\bf Theorem 1.} {\it All the rotation linear W-symmetries of an
Ito equation are also symmetries of the associated Stratonovich
equation, and viceversa. Dilation W-symmetries of an Ito equation
are also symmetries of the associated Stratonovich equation (and
vice versa) if and only if the diffusion matrix is spatially
constant.}

\medskip\noindent
{\bf Corollary 1.} {\it If the diffusion matrix $\s^i_{\ k}$ in
\eqref{eq:Ito} is constant w.r.t. space variables, then all
W-symmetries of the Ito equation are also symmetries of the
corresponding Stratonovich equation.}
\bigskip

It is also a simple consequence of the above Theorem 1 that
Proposition 1 extends to this kind of symmetries:

\medskip\noindent
{\bf Theorem 2.} {\it Rotation linear W-symmetries of an Ito
equation \eqref{eq:Ito} are preserved under changes of variables
defined by an admissible W-map $x^i = \Phi^i (y,t;z)$, $w^k =
R^k_{\ m} z^m$. The same applies to all linear W-symmetries if the
diffusion matrix is constant w.r.t. space variables.}

\medskip\noindent
{\bf Remark 18.} We stress that the limitation (to linear
structure) only regards the symmetry vector field, while -- as
clear from our previous discussion -- it does not affect the form
of the considered W-map. On the other hand, in practice we will
consider W-maps associated to W-symmetries, and as these are
linear (we recall that here ``linear'' only refers to the $w$
component of the symmetry and map) the W-maps will also be linear,
at the exception of dilations. \EOR

\medskip\noindent
{\bf Remark 19.} As stressed above, see Remarks {\RK} and {\RKK},
in some cases one may wish to consider equations more general than
Ito ones. Similarly, one may wish to consider maps such that the
random processes underlying the $x$ stochastic processes are
allowed to be more general than Wiener ones. This is why we have
taken the seemingly odd choice of performing a part of our
computations considering general $h (x,t;w)$ functions, albeit in
the present paper we are only interested in the linear case $h = R
z$. \EOR

\subsection{Examples}
\label{sec:exa}

We will now consider some explicit Examples illustrating our
results and in particular Theorem 1. As we deal with
time-autonomous equations, we will consider time-independent
symmetries, thus slightly simplifying the discussion and the
(intermediate) explicit formulas.

\medskip\noindent
{\bf Example 3.} We consider the scalar Ito equation \beq d x \ =
\ \la x \, d t \ + \ \mu \, d w \ , \eeq with $\la$ and $\mu$
nonzero real constants. In this case (as always for a constant
diffusion coefficient) the associated Stratonovich equation reads
just in the same way, i.e. $b(x,t) = f(x,t) = \la x$.

As for the determining equations, those for the Ito equation, i.e.
\eqref{eq:WIde1} and \eqref{eq:WIde2}, read
$$ \la \, x \, \vphi_x \ - \ \la \, \vphi \ + \ \frac12
\Delta(\vphi) \ = \ 0 \ , \ \ \vphi_w \ + \ \mu \, \vphi_x \ = \
\mu \, R \ ; $$ while those for the associated Stratonovich
equation, i.e. \eqref{eq:WSde1} and \eqref{eq:WSde2}, read
$$ \la \, x \, \vphi_x \ - \ \la \, \vphi \ = \ 0 \ , \ \ \vphi_w \ + \ \mu \, \vphi_x \ = \
\mu \, R \ . $$ Thus the second equation in the two sets is the
same (as always), while the first ones are different. However,
when we restrict to solutions of the second equation, i.e. to
$$ \vphi \ = \ R \, x \ + \ \Theta \[ \zeta \] \ , \ \ \ \
\zeta := w - x/\mu \ , $$ the two
equations coincide. Hence the symmetries of the Ito and of the
associated Stratonovich equation coincide, as stated by our
Theorem 1.

Actually, one finds immediately that enforcing also the first
equation requires $\Theta = 0$, thus the symmetries reduce to the
obvious scaling one, $(x,w) \to (s x , s w)$. This one-parameter
group ($s \in {\bf R}_+$) is generated by the vector field
$$ X \ = \ x \, \pa_x \ + \ w \, \pa_w \ . $$
Note that here we have $\phi = x$, $R=1$. \EOR

\medskip\noindent
{\bf Example 4.} Consider more generally the scalar Ito equation
\beq d x \ = \ \la \, x \ d t \ + \ \mu \, x^\a \ d w \ ; \eeq
here again $\la, \mu$ are nonzero real constants. For $\a = 0$
this reduces to the previous Example, so we assume the real
constant $\a$ is also nonzero.

It is clear that this equation is invariant under the scalings
$(x,w) \to (s x , s^{1 - \a} w)$, generated by the vector field
$$ X \ = \ x \, \pa_x \ + \ (1 - \a) \, \pa_w \ ; $$
note the case $\a = 1$ is not of interest here, as it does not
correspond to a W-symmetry.

In this case the associated Stratonovich equation is \beq d x \ =
\ \( \la x \ - \ \frac12 \, \a \, \mu^2 \, x^{(2 \a - 1)} \) \ d t
\ + \ \mu \, x^\a \ d w \ . \eeq It is quite obvious that this
equation is invariant under the scaling mentioned above if and
only if $2 \a - 1 = 1$, i.e. for the ``uninteresting case'' $\a =
1$.

The second equation in the set of determining ones is, in both
cases, \beq \vphi_w \ + \ \mu \, x^\a \vphi_x \ - \ \a \, \mu \,
x^{\a-1} \, \vphi \ = \ \mu \, x^\a \, R \ . \eeq The most general
solution to this equation is
$$ \vphi (x,w) \ = \ \frac{R}{\a - 1} \, x \ + \ \Theta \[ \frac{x
+ (\a - 1) \mu x^\a w}{(\a - 1) \mu x^\a } \] \ , $$ with $\Theta$
an arbitrary function. When we consider \eqref{eq:WIde1} we obtain
that $\Theta$ must be zero. We are thus left with vector fields of
the form $X = [R/(\a-1)] (x \pa_x + (\a-1) w \pa_w )$; we can of
course choose $R= \a-1$, hence $\vphi = x$, and we are left with
the symmetry generator \beq  X \ = \ x \, \pa_x \ + \ (\a - 1 ) w
\, \pa_w \ . \eeq

Direct substitution in \eqref{eq:WSde1} shows that (unless $\a =
1$) this is \emph{not} a symmetry for the associated Stratonovich
equation. In fact, that equation reduces now to the identity $$ \a
\, (\a - 1) \, \mu^2 \ x^{2 \a - 1} \ = \ 0 \ , $$ which is
satisfied only in the cases we have excluded ($\mu = 0$, $\a=0$,
$\a=1$). \EOR

\medskip\noindent
{\bf Example 5.} In the one-dimensional case one will most
frequently find  scaling symmetries, if any, but it is possible to
build some (admittedly, rather artificial) example which admits a
nonlinear $\vphi (x)$. To this aim, consider the equation
$$ d x \ = \ f(x) \, d t \ + \ \s (x) \, d w $$
with the functions
\begin{eqnarray*}
f(x) &:=& c_1 \, x^2 \ + \ x^2 \ \( x \, \exp[ 2 /
x] \ - \ 2  \ \mathtt{Ei}[2 / x] \) \ , \\
\s (x) &:=& c_1 \, x^2 \ \exp[1/x] \ , \end{eqnarray*} with
$\mathtt{Ei} (z)$ denoting the exponential integral function
$$ \mathtt{Ei} (z) \ = \ - \, \int_{- z}^\infty \frac{e^{- t}}{t}
\, d t $$ (the principal value is taken here).

In this case we have (only) the W-symmetry vector field
$$ X \ = \ x^2 \, \pa_x \ + \ w \, \pa_w \ . $$
Obviously this example was built by reverse engineering, i.e.
assigning $\vphi,R$ and looking at the determining equations as
equations for $\vphi , \s$.

One can check that this $X$ is not a symmetry for the associated
Stratonovich equation. \EOR

\medskip\noindent
{\bf Example 6.} We will now consider a multi-dimensional
generalization of Example 3, i.e. the stochastic linear oscillator
(no sum on $i$ in this Example) \beq d x^i \ = \ \la_i \, x^i \ d
t \ + \ \mu_i \, d w^i \ \ \ (i=1,...,n) \ . \eeq

It is clear that this will have scaling symmetries, and in the
isotropic case $\la_i = \la$, $\mu_i = \mu$ also rotation
symmetries (with partial rotation symmetries in case of partially
isotropic oscillator). In this Example we will not assume any
relation between the constants; in the next Example we will
consider the isotropic linear oscillator.

Let us discuss in detail the case $n=2$, assuming all diffusion
constants $\mu_i$ appearing in the system are nonzero; the general
case would be not too different. We will write all indices as
lower ones for typographical convenience and in order to avoid any
possible confusion.

In this $n=2$ case the second set of determining equations (common
to the Ito and the Stratonovich case) is solved by the method of
characteristics and yields the general solutions
\begin{eqnarray}
\vphi_1 &=& R_{1 1} \ x_1 \ + \ \frac{\mu_1}{\mu_2} \
R_{1 2} \ x_2 \ + \ \psi_1 (z_1,z_2) \ , \\
\vphi_2 &=& \frac{\mu_2}{\mu_1} R_{2 1} \ x_1 \ + \ R_{2 2} \ x_2
\ + \ \psi_2 (z_1,z_2) \ , \end{eqnarray} where we have written
$$ z_1 \ := \ w_1 \ - \ \frac{x_1}{\mu_1} \ , \ \ \
z_2 \ := \ w_2 \ - \ \frac{x_2}{\mu_2} \ , $$ and $\psi_i$ are
arbitrary smooth functions of their arguments $(z_1,z_2)$.

Now the first set of determining equations for the Ito equations
reads
\begin{eqnarray}
\la_1 \, \psi_1 & + & \( \frac{\la_1 (\pa \psi_1 / \pa
z_1)}{\mu_1} \) \, x_1 \nonumber \\
& & \ + \ \( \frac{\la_1 \mu_1^2 R_{12} \, - \, \la_2 \mu_1^2
R_{12} \, + \, \la_2 \mu_1 (\pa \psi_1 / \pa
z_2)}{\mu_1 \, \mu_2} \) \, x_2 \ = \  0 \ , \label{eq:RE31} \\
\la_2 \, \psi_2 & - & \( \frac{(\la_1 - \la_2) \mu_2 R_{21} -
\la_1 (\pa \psi_2 / \pa z_1)}{\mu_1} \) \, x_1 \nonumber \\
 & & \ + \ \(
\frac{\la_2 (\pa \psi_2 / \pa z_2)}{\mu_2} \) \, x_2 \ = \  0 \ .
\label{eq:RE32}
\end{eqnarray}

These two (uncoupled) equations are again solved by the method of
characteristics. Recalling that $\psi_i = \psi_i (z_1,z_2)$ we
readily get that for $\la_1 \not= 0 \not= \la_2$ we necessarily
have $\psi_1 = 0 = \psi_2$. With these, we are reduced to
\begin{eqnarray}
\frac{\mu_2}{\mu_1} \ (\la_2 - \la_1 ) \ R_{12} &=& 0 \ , \\
\frac{\mu_2}{\mu_1} \ (\la_1 - \la_2 ) \ R_{21} &=& 0 \ ;
\end{eqnarray}
thus in the case $\la_1 \not= \la_2$ we necessarily have
$R_{12}=R_{21}=0$. Finally, the coefficient of the symmetry vector
field are
$$ \vphi^1 \ = \ R_{11} \, x_1 \ , \ \ \vphi^2 \ = \ R_{22} \, x_2
\ . $$ Thus we have two scaling symmetries:
$$ Y_1 \ = \ x_1 \, \pa_1 \ + \ w_1 \, \^\pa_1 \ , \ \ Y_2 \ = \  x_2
\, \pa_2 \ + \ w_2 \, \^\pa_2 \ . $$ It is more convenient to
consider the sum and difference of these two, providing
\begin{eqnarray}
X_1 &=& Y_1 + Y_2 \ = \ x_1 \, \pa_1 \ + \ x_2 \, \pa_2
\ + \ w_1 \, \^\pa_1 \ + \ w_2 \^\pa_2 \ , \\
X_2 &=& Y_1 - Y_2 \ = \ x_1 \, \pa_1 \ - \ x_2 \, \pa_2
\ + \ w_1 \, \^\pa_1 \ - \ w_2 \^\pa_2 \ , \\
\ . \end{eqnarray} The $R$ matrices associated to these are
respectively
$$ R_1 \ = \ \begin{pmatrix} 1 & 0 \\ 0 & 1 \end{pmatrix} \ , \ \
\ R_2 \ = \ \begin{pmatrix} 1 & 0 \\ 0 & - 1 \end{pmatrix} \ . $$
The first one generates scalings $(w_1 ,w_2) \mapsto (s w_1 , s
w_2)$ and hence conformal maps, while the second one generates the
one parameter group $(w_1,w_2) \mapsto (s w_1 , s^{-1} w_2 )$ and
hence does not correspond to conformal maps.

Thus $X_1$ is an acceptable W-symmetry generator, while $X_2$
fails to preserve the independence of the Wiener processes and is
therefore \emph{not} an acceptable W-symmetry generator.

One can check by explicit computation that $X_1,X_2$ (or more
precisely the corresponding vectors $\vphi$ and matrices $R$) also
satisfy the determining equations for symmetries of the associated
Stratonovich equation. \EOR

\medskip\noindent
{\bf Example 7.} We will now consider the \emph{isotropic}
stochastic linear oscillator \beq d x^i \ = \ \la \, x^i \ d t \ +
\ \mu \, d w^i \ \ \ (i=1,...,n) \ . \eeq

Now after solving the common set of determining equations we have
(we write again all indices as lower ones to avoid confusion) \beq
\vphi_1 \ = \ \psi_1 \ + \ R_{11} \, x_1 \ + \ R_{12} \, x_2 \ , \
\ \ \vphi_2 \ = \ \psi_2 \ + \ R_{21} x_1 \ + \ R_{22} \, x_2 \ .
\eeq Plugging this into the equations \eqref{eq:WIde1},
\eqref{eq:WIde2} we obtain again that $\psi_i = 0$, but now the
final result is that \beq \vphi_1 \ = \ R_{11} \, x_1 \ + \ R_{22}
\, x_2 \ , \ \ \ \vphi_2 \ = \ R_{21} x_1 \ + \ R_{22} \, x_2 \ .
\eeq

This leaves us with four symmetry vector fields,
\begin{eqnarray*}
Y_1 &=& x_1 \, \pa_1 \ + \ w_1 \, \^\pa_1 \ , \ \
Y_2 \ = \ x_2 \, \pa_2 \ + \ w_2 \, \^\pa_2 \ ; \\
Y_3 &=& x_2 \, \pa_1 \ + \ w_2 \, \^\pa_1 \ , \ \
 Y_4 \ = \  x_1 \, \pa_2 \ + \ w_1 \, \^\pa_2 \ .
\end{eqnarray*}

Again it is more convenient to consider sum and differences of
these, i.e.
\begin{eqnarray*}
X_1 &=& Y_1 \, + \, Y_2 \ = \ x_1 \, \pa_1 \ + \ x_2 \, \pa_2 \ +
\ w_1 \, \^\pa_1 \ + \ w_2 \, \^\pa_2 \ , \\
X_2 &=& Y_1 \, - \, Y_2 \ = \ x_1 \, \pa_1 \ - \ x_2 \, \pa_2 \ +
\ w_1 \, \^\pa_1 \ - \ w_2 \, \^\pa_2 \ ; \\
X_3 &=& Y_3 \, + \, Y_4 \ = \ x_2 \, \pa_1 \ + \ x_1 \, \pa_2 \ +
\ w_2 \, \^\pa_1 \ + \ w_1 \, \^\pa_2 \ , \\
X_4 &=& Y_4 \, - \, Y_4 \ = \ x_2 \, \pa_1 \ - \ x_1 \, \pa_2 \ +
\ w_2 \, \^\pa_1 \ - \ w_1 \, \^\pa_2 \ . \end{eqnarray*}

The first two are the scaling symmetries always present and
discussed in the general case, while $X_4$ generates equal
rotations in the $(x_1,x_2)$ and in the $(w_1,w_2)$ planes, and
$X_3$ generates equal \emph{hyperbolic} rotations in the
$(x_1,x_2)$ and in the $(w_1,w_2)$ planes.

It is immediate to check that $X_1$ and $X_4$ generates groups of
conformal transformations (in particular, in the $(w_1,w_2)$
plane), while this is not the case for $X_2$ and $X_3$.

Thus in view of our general discussion only $X_1$ and $X_4$ are
acceptable W-symmetry generators.

We note that the Lie algebra structure of the $X_i$ fields is as
in the following commutator table, where as usual the entry
$(i,j)$ represents $[X_i,X_j]$:
$$ \begin{tabular}{|c||c|c|c|c||}
\hline
    & $X_1$ & $X_2$    & $X_3$     & $X_4$ \\
\hline
$X_1$ &  0  &  0     &  0      &  0  \\
$X_2$ &  0  &  0     & $- 2 X_4$ & $- 2 X_3$ \\
$X_3$ &  0  & $2 X_4$  &  0      & $2 X_2$ \\
$X_4$ &  0  & $2 X_3$  & $- 2 X_2$ &  0 \\
\hline
\end{tabular} $$
Note that the acceptable W-symmetries $\{ X_1 , X_4 \}$ span a Lie
subalgebra, as they should (moreover, in the case under study this
is Abelian).

Again the symmetry vector fields are also symmetries for the
associated Stratonovich equation. \EOR

\medskip\noindent
{\bf Example 8.} We will now consider a generalization of Example
7 with non constant diffusion matrix. We deal with the isotropic
stochastic non-linear oscillator \beq d x^i \ = \ \a (|x|^2) \,
x^i \ d t \ + \ \b (|x|^2) \, d w^i \ \ \ (i=1,...,n) \ ; \eeq in
general (that is, unless $\a$ and $\b$ are actually both constant
functions) this has no scaling symmetries, but retains rotational
symmetries. Again we just consider the case $n=2$, so that now
rotations are generated by the single vector field \beql{eq:XRot}
X \ = \ - x^2 \, \pa_1 \ + \ x^1 \, \pa_2 \ - \ w^2 \, \^\pa_1 \ +
\ w^1 \, \^\pa_2 \ = \ J^i_{\ k} \, \( x^k \, \pa_i \ + \ w^k \,
\^\pa_i \) \ , \eeq with the same notation as above.

Here we will not look for the most general solution to the
determining equations, but just note that -- as can be checked by
direct computation, the $X$ above is an acceptable W-asymmetry
generator. In fact, choosing
$$ \vphi \ = \ \begin{pmatrix} - x_2 \\ x_1 \end{pmatrix} \ , \ \
R \ = \ \begin{pmatrix} 0 & - 1 \\ 1 & 0 \end{pmatrix} $$ the Ito
determining equations \eqref{eq:WIde1}, \eqref{eq:WIde2} are
satisfied.

One can also check, again by direct computation, that in this case
the Stratonovich determining equations \eqref{eq:WSde1},
\eqref{eq:WSde2} are also satisfied. Note that in this case the
diffusion matrix $$ \s \ = \ \begin{pmatrix} \mu [ x_1^2 + x_2^2]
& 0 \\ 0 & \mu [ x_1^2 + x_2^2] \end{pmatrix} $$ is non constant,
but $R$ is skew-symmetric.

The same result is obtained if only one of the two function $\a$,
$\b$ is non constant, as can be checked by explicit computations
(in these cases determining the most general solution to the
determining equations is rather simple, and one finds indeed only
the rotations given above). \EOR

\medskip\noindent {\bf Example 9.} It may be interesting to look
again at the linear isotropic stochastic oscillator \beql{eq:exa1}
d x^i \ = \ - \, K \ x^i \, d t \ + \ \s \, d w^i \eeq with $K$
and $\s$ real constants in arbitrary dimension ($i = 1,...,n$),
but using the ``general'' notation $h^k$ for the $\^\pa_k$
component of the symmetry vector field. Then the Ito determining
equations read
\begin{eqnarray*}
\pa_t (\vphi^i - \s h^i) &=& K \, \sum_{j=1}^n \pa_j (\vphi^i -
\s h^i) \ - \ \frac12 \De (\vphi^i - \s h^i ) \\
\^\pa_k (\vphi^i - \s h^i) &=& - \, \s \, \pa_k (\vphi^i - \s h^i
) \ . \end{eqnarray*}

Thus in this case we can pass to consider $\psi^i := \vphi^i - \s
h^i$, hence be reduced to considering a single set of functions
$\psi^i (x,t;w)$, as for standard (in general, random) symmetries.
Doing this we obtain the determining equations
\begin{eqnarray*}
\pa_t \psi^i &=& K \, \sum_{j=1}^n \pa_j \psi^i \ - \ \frac12 \De \psi^i \\
\^\pa_k \psi^i &=& - \, \s \, \pa_k \psi^i \ , \end{eqnarray*}
which are just the determining equations for standard symmetries
of \eqref{eq:exa1}. \EOR

\medskip\noindent {\bf Example 10.} We have seen that isotropic
nonlinear stochastic oscillators admit (only) rotation symmetries;
in particular, in the two-dimensional case we have the single
W.symmetry vector field \eqref{eq:XRot}. We can look at the
inverse problem, i.e. identifying Ito equations admitting rotation
symmetries; we will again confine ourselves to the two-dimensional
case.

This just requires to look at the determining equations with
$\vphi$ and $h$ (i.e. $R$) assigned, and $f^i$, $\s^i_{\ k}$ as
unknown.

In this case, writing $r^2 = x_1^2+x_2^2$, \eqref{eq:WIde2} yield
$$ \s \ = \ \begin{pmatrix} \a (r^2,t) & - \b (r^2,t) \\
\b (r^2,t) & \a (r^2 ,t ) \end{pmatrix} \ , $$ with $\a$ and $\b$
smooth functions of their arguments. As for \eqref{eq:WIde1},
these yield
$$ f^i \ = \ G (r^2,t) \ x^i \ . $$
These are also the most general solutions to \eqref{eq:WSde1},
\eqref{eq:WSde2}. \EOR

\medskip\noindent {\bf Example 11.} All Examples considered so far
yielded split W-symmetries; one could wonder if non-split
W-symmetries are possible at all. The answer is positive, as shown
by the trivial case of a constant coefficients scalar Ito equation
\beql{eq:nosplit} d x \ = \ A \, d t \ + \ B \, d w \ ; \eeq note
we must assume $B \not= 0$, or we would not have a stochastic
equation.

Now the determining equations are
\begin{eqnarray*}
\vphi_t &+& A \, \vphi_x \ + \ \frac12 \( \vphi_{ww} + 2 B
\vphi_{xw} + B^2 \vphi_{xx} \) \ = \ 0 \ , \\
\vphi_w &+& B \, \vphi_x \ - \ B \, R \ = \ 0 \ . \end{eqnarray*}
The second equations yields immediately
$$ \vphi \ = \ R \, x \ + \ \psi (z , t) \ , \ \ \ z := w - x/B \
. $$ Plugging this into the first determining equation we get
$$ \psi_t \ - \ (A/B) \, \psi_z \ + \ (A/B) R \ = \ 0 \ , $$
which in turn yields
$$ \psi (z,t) \ = \ A \, R \, t \ + \ \Theta (\zeta) \ , \ \ \
\zeta = z + (A/B) t \ . $$ Thus, we always have the W-symmetries
$$ X_\Theta \ = \ \[ x - A t + \Theta (\zeta) \] \, \pa_x \ + \ \pa_w \
, $$ which is a non-split W-symmetry provided $\Theta \not= 0$; in
particular, with the choice $\Theta (y) = B R y$ we get
$$ X \ = \ (B \, w) \ \pa_x \ + \ \pa_w  $$
i.e. a time-autonomous non-split nontrivial W-symmetry. (In
Appendix B it will be shown that this is essentially the only
example of non-split W-symmetries for scalar Ito equations.)

Note that \eqref{eq:nosplit} is solved as $$ x(t) \ = \ x(t_0) \ +
\ A \, (t - t_0) \ + \ B \, [ w(t) - w(t_0)] \ , $$ so it is
immediate to check that (the one-parameter group generated by) $X$
maps solutions into solutions. \EOR


\section{Application of W-symmetries}
\label{sec:Wkoz}

The general idea behind the use of symmetries to simplify and/or
solve differential equations (deterministic or stochastic) is to
pass to \emph{symmetry-adapted coordinates}.

This is also the case for Kozlov theorems, discussed in Sections
\ref{sec:Koz} and \ref{sec:Kozsyst}; in fact, in this case one
change coordinates so that the symmetry vector field is
transformed into a vector field along one of the new coordinates,
and the independence of the equation on this allows for a direct
(partial, for systems) integration.

We will thus try to follow the same approach here. It will be
quite clear, even from the simplest example of stochastic
oscillators, that the outcome will be quite different from that
seen in the case of standard (deterministic or stochastic)
symmetries.

\subsection{Scalar equations}

The problem is already apparent if we consider one-dimensional
systems, i.e. scalar equations. We will just cons8der autonomous
equations, thus time will not need to be considered even in the
functional dependencies (this will just simplify our notation,
with no loss of generality, as the reader can easily check).

In our case of W-symmetries, the standard Kozlov change of
coordinates \beq \xi \ = \ \int \frac{1}{\vphi (x,t,w)} d x \eeq
(which is guaranteed to map the Ito equation into an Ito equation)
does not suffice to rectify the vector field $$ X \ = \ \vphi
(x,t,w) \, \pa_x \ + \ R w \, \pa_w \ , $$ and hence guarantee
integrability. In fact, now
$$ X (\xi) \ = \ \vphi \, \frac{1}{\vphi} \ + \ R w \int \left(
    \frac{\pa}{\pa w} \, \frac{1}{\vphi} \right) \, d x \ = \ 1
    \ - \ R \, w \ \int \frac{\vphi_w}{\vphi^2} \, d x \ , $$ and
as the second term in general is not zero, we do not have $X =
\pa_\xi$, hence the r.h.s. of the transformed equation is not
independent of $\xi$ and it cannot be explicitly integrated.

This is already apparent when considering Stratonovich equations,
i.e. is not related to problems arising from applying the Ito rule
when changing coordinates.

This is not surprising: as $X$ has also a component along $\pa_w$,
we should also change the $w$ variable, i.e. pass from $(x,w)$ to
$(\xi (x,w), \zeta(x,w))$ variables in order to  have $X =
\pa_\xi$ and hence guarantee direct integration of the equation $d
\xi = F dt + S d\zeta$ for $\xi$.

\medskip\noindent
{\bf Example 12.} Consider the Stratonovich equation (linear
stochastic oscillator) \beql{eq:solK0} d x \ = \ \la \, x \ dt \ +
\ \mu \circ dw \ , \eeq with $\la , \mu$ real constants. This
admits as symmetry generator the scaling vector field
$$ X \ = \ x \, \pa_x \ + \ w \, \pa_w \ . $$ The Kozlov change of
variable is then \beq \xi \ = \ \int \frac{1}{\vphi} d x \ = \
\int \frac{1}{x} \, d x \ = \ \log x \ ; \ \ x = e^\xi \ . \eeq In
terms of this variable, we have of course $$ d \xi \ = \
\frac{1}{x} \ d x \ = \ \frac{1}{x} \ \left[ \la x d t \ + \ \mu d
w \right] \ , $$ so our original equation \eqref{eq:solK0} reads
now \beql{eq:solK1} d \xi \ = \ \la \, d t \ + \ \mu \, e^{- \xi}
\circ d w \ . \eeq

The vector field does now read $$ X \ = \ \pa_\xi \ + \ w \, \pa_w
\ , $$ and it is immediate to check this is indeed a symmetry of
\eqref{eq:solK1}. The problem is that \eqref{eq:solK1} can
\emph{not} be directly integrated.

As shown by the fact we are considering a Stratonovich equation,
this is not even related to the Ito rule, but to the very nature
of W-symmetries. \EOR

\subsection{Adapted variables}

As hinted above, our strategy in using W-symmetries should be
equal in principles, but slightly different in practice, to the
one for standard symmetries. That is, we should pass from the old
variables $(x^i,w^k)$ ($i = 1,...,n$, $k = 1,...,m$; possibly with
$m=n$) to new coordinates \beql{eq:Kcv} \xi^i (x,t;w) \ , \ \
\zeta^k (x,t;w) \eeq such that in the new variables the symmetry
vector field $X$ -- which we assume to be of the type identified
in Sect.\ref{sec:Wsymm}, see \eqref{eq:WmapKX} -- reads
$$ X \ = \ \frac{\pa}{\pa \xi^n} \ . $$

Now the equations will read as \beql{eq:SK0} d \xi^i \ = \ F^i \,
d t \ + \ S^i_{\ k} \, d \zeta^k \ \ \ \ (i=1,...,n) \ ; \eeq as
$X$ is a symmetry, we will have \beq \frac{\pa F^i}{\pa \xi^n} \ =
\ 0 \ = \ \frac{\pa \S^i_{\ k}}{\pa \xi^n} \ . \eeq

Thus, \emph{if} we are able to solve the reduced system
\beql{eq:SK1} d \xi^i \ = \ F^i \, d t \ + \ S^i_{\ k} \, d
\zeta^k \ \ \ \ (i=1,...,n-1) \ , \eeq then the solution to the
last equation \beql{eq:SK2} d \xi^n \ = \ F^n \, d t \ + \ S^n_{\
k} \, d \zeta^k  \eeq amounts to a direct (stochastic)
integration.

This is only apparently identical to what happens for standard
symmetries. Actually, a substantial difference arises now due to
the more general form of the change of variables \eqref{eq:Kcv}.

In fact, now
\begin{enumerate}
\item The functions $F^i$ and $S^i_{\ k}$ appearing in
\eqref{eq:SK0} will in general depend not only on the $(\xi,t)$
variables but \emph{also} on the $\zeta^k$; \item The $\zeta^k$
will in general \emph{not} be Wiener processes.
\end{enumerate}

Each of these features makes that the new equation \eqref{eq:SK0}
is \emph{not} of Ito type. As remarked above, see Remarks \RK and
{\RKK}, this in itself is not forbidding the reduced equation can
be integrated and thus the W-symmetry reduction procedure
maintains some interest (see also Section \ref{sec:stochosc}
below).

On the other hand, it should be noted that in the case of multiple
symmetries we get out of what is covered by the presently existing
theory. In fact, one can very well consider stochastic
differential equations which are not of Ito (or Stratonovich) type
\cite{Kampen}; but as soon as we deal with an equation which is
not of Ito type, we cannot use the correspondence with the
Stratonovich form in order to guarantee that symmetries will
survive a change of variables\footnote{Albeit one would expect
this to be the case, at least for admissible symmetries.}, hence
we cannot -- at the present stage of our mathematical knowledge --
ignite the recursion procedure which was able to guarantee
multiple reduction for standard symmetries, i.e. in the frame of
standard Kozlov theory \cite{GL2}.\footnote{Needless to say, this
is not a ``no go'' result, but rather calls for a study of a more
general framework for the use of symmetry in the stochastic
realm.}

We summarize or discussion as a formal statement\footnote{We
recall that the adapted variables $(y,z)$ for a vector field $X$
are those such that $X = (\pa / \pa y^n)$.}, which will then be
illustrated by studying stochastic oscillators in the next
subsection.

\medskip\noindent
{\bf Theorem 3.} {\it Let the Ito equation \eqref{eq:Ito} admit a
nontrivial W-symmetry with generator $X$. Passing to adapted
variables $(y,z)$ the equation is mapped into a system of
stochastic differential equations \beql{eq:thm3} d y^i \ = \ F^i
\, d t \ + \ S^i_{\ k} \, d z^k \eeq with
$$ \frac{\pa F^i}{\pa y^n} \ = \ 0 \ = \ \frac{\pa S^i_{\ k}}{\pa
y^n} $$ for all $i,k = 1,...,n$; these are in general not of Ito
type, i.e. the coefficients $F^i$ and $S^i_{\ k}$ can depend on
the driving stochastic processes $z^k$.}

\medskip\noindent
{\bf Corollary 2.} {\it If the $n$-dimensional system of Ito
equations \eqref{eq:Ito} admits a nontrivial W-symmetry, it can be
mapped into a system of stochastic differential equations
\eqref{eq:thm3} which decouples into an autonomous systems of
$(n-1)$ equations plus a ``reconstruction equation'' \beq d y^n \
= \ F^n [ y^1 , ... , y^{n-1} ; z^1,...,z^n] \, d t \ + \ S^n_{\
k} [ y^1 , ... , y^{n-1} ; z^1,...,z^n] \, d z^k \ . \eeq}

\subsection{Example. Stochastic oscillators}
\label{sec:stochosc}

We will now apply the previous discussion to the simple but
relevant case of \emph{stochastic oscillators}, considering both
dilation (scaling) and rotation symmetries. We will confine
ourselves to the simplest cases, i.e. those in one and two spatial
dimensions; these were already considered in the Examples of
Section \ref{sec:exa}, so we will build on the computations
performed there.

\subsubsection{Scaling}

We start by considering the linear stochastic oscillator (in one
dimension, as this will suffice to point out the problem we need
to discuss) \beql{eq:sol} d x \ = \ \la \, x \, d t \ + \ \mu \, d
w \ ; \eeq here $\la$ and $\mu$ are real constants. As discussed
above (see Section \ref{sec:exa}) eq. \eqref{eq:sol} admits the
simple scaling symmetry generator \beq X \ = \ x \, \pa_x \ + \ w
\, \pa_ w \ ; \eeq which generates the one-parameter group of
scalings $ x \to s x$,  $w \to s w$.

The invariant quantity under this is $ \zeta = w/x$, and the
vector field satisfies $X(\xi ) = 1$ e.g. for $\xi = \log (x)$. We
will thus pass to coordinates
$$ \xi = \log (x) \ , \ \ \zeta = w/x \ ; $$ the inverse
change of variables is
$$ x = e^\xi \ , \ \ w = e^\xi \ \zeta \ . $$
In these coordinates, the symmetry vector field reads simply
$$ X \ = \ \pa_\xi \ . $$

The equation \eqref{eq:sol} will now be written as \beql{eq:sol2}
d \xi \ = \ F \, d t \ + \ S \, d \zeta \ , \eeq with $F$ and $S$
functions which will now be determined.

Using the Ito rule we have\footnote{The computation is performed
here following the scheme suggested in \cite{GL2}; other ways of
performing the same computation are of course also possible.}
\begin{eqnarray*}
d x &=& e^\xi \, d \xi \ + \ \frac12 \left( e^\xi \, S^2 \right)
\, d t \\
d w &=& e^\xi \, \zeta \, d \xi \ + \ e^\xi \, d \zeta \ + \
\frac12 \left( 2 S e^\xi \, + \, S^2 e^xi \zeta \right) \, d t \ .
\end{eqnarray*}
Thus \eqref{eq:sol} reads now, with simple algebra,
$$ d \xi \ = \ \left[ \frac{\la + \mu S (1 + S \zeta ) - S^2 / 2
}{1 - \mu \zeta } \right] \ dt \ + \ \left( \frac{\mu}{1 - \mu
\zeta } \right) \ d \zeta \ . $$ This shows that
$$ S \ = \ \left( \frac{\mu}{1 - \mu
\zeta  } \right) $$ and inserting this into the coefficient of
$dt$ we obtain that
$$ F \ = \ \left[ \la \ + \ \frac12 \, \frac{\mu^2}{(1 - \mu \zeta
)^2 } \right] \ \left( \frac{1}{1 - \mu \zeta } \right) \ . $$ The
explicit expressions of $F$ and $S$ are not relevant; the
important thing are their functional dependencies. That is,
eq.\eqref{eq:sol2} is more precisely rewritten as \beql{eq:sol3} d
\xi \ = \ F (\zeta) \, d t \ + \ S(\zeta) \, d \eta \ . \eeq

We conclude that:
\begin{itemize}
\item The vector field $X$ is still a symmetry of the transformed
equation, as stated in our Theorem 2; \item But the transformed
equation is \emph{not} of Ito type, as the coefficients depend
explicitly on the driving random process $\zeta$. \end{itemize}

This situation should be compared with that seen above in Example
2, see also Remark \RK. Albeit the equation is not of Ito type, we
immediately have
$$ \xi (t) \ = \ \int F [\zeta (t)] \, d t \ + \ \int S[\zeta (t)]
\, d \zeta (t) $$ and $\xi (t)$ is recovered by a stochastic integral.

Note that $\zeta (t)$ is in general \emph{not} a Wiener process,
as clear from the transformation law linking $\zeta (t) =
w(t)/x(t)$ to the Wiener process $w(t)$.

Once we have determined $\xi (t)$ for a given realization of the
stochastic process $\zeta (t)$, the $x(t)$ is immediately
recovered as $$ x  (t) \ = \ \exp [ \xi (t) ] \ . $$

We proceed exactly in the same way, apart from introducing some
indices, in considering multi-dimensional linear stochastic
oscillators and their scaling symmetries.

\subsubsection{Rotation}

The same qualitative situation is found if we work in higher
dimensions and consider an isotropic stochastic oscillator (in
this case, linear or nonlinear) and its rotational symmetry.

Consider, for the sake of definiteness, the two-dimensional
setting (i.e. $n=2$; we write again all indices as lower ones to
avoid any confusion) for the general equation \eqref{eq:WROT}. In
this case the W-symmetry generator is \beql{eq:XW} X \ = \ x_2
\pa_1 - x_1 \pa_2 + w_2 \^\pa_1 - w_1 \^\pa_2 \ . \eeq

We now want to consider adapted coordinates; in this case they are
polar coordinates in both the $x$ and the $w$ space, i.e.
$$ r = \sqrt{x_1^2 + x_2^2} , \ \vartheta = \arctan(x_2/x_1) ; \ z
= \sqrt{w_1^2 + w_2^2} \ , \ \xi = \arctan(w_2/w_1) \ . $$ This
corresponds, obviously, to
$$ x_1 = r \cos (\vartheta) , \ x_2 = r \sin (\vartheta) ; \ w_1 =
z \, \cos (\xi ) , \ w_2 = z \, \sin (\xi ) \ . $$ The vector
field \eqref{eq:XW} reads now $X = \pa_\vartheta + \pa_\xi$.

With a standard application of Ito rule, we have
\begin{eqnarray*}
d r &=& \frac{1}{2r} \( S^2 + 2 r^2 F \) \, d t \ + \ 2 S r \[
\cos (\vartheta - \xi) \, d z \ + \ z \sin(\vartheta - \xi) \, d
\xi \] \ , \\
d \vartheta &=& \frac{S z}{r} \cos (\vartheta - \xi) \, d \xi \ -
\ \frac{1}{r} \sin (\vartheta - \xi ) \, d z \ . \end{eqnarray*}
It is apparent that these are invariant under the simultaneous
rotations $\vartheta \to \vartheta \ \delta$, $\xi \to \xi +
\delta$; that is, under $X$.

With a further trivial change of variables, i.e. switching from
$\vartheta $ to $\psi = \vartheta - \xi$, these become
\begin{eqnarray}
d r &=& \frac{1}{2r} \( S^2 + 2 r^2 F \) \, d t \ + \ 2 S r \[
\cos (\psi) \, d z \ + \ z \sin(\psi) \, d
\xi \] \ , \nonumber \\
d \psi &=& \( \frac{S}{r} z \cos \psi \ - \ 1 \) \, d \xi \ - \
\frac{1}{r} \, \sin \psi \, d z \ . \label{eq:exa13}
\end{eqnarray} (These are immediately seen to be invariant under a
shift in $\xi$, i.e. to admit the W-symmetry $X_0 = \pa/\pa \xi$;
but this is not admissible in view of the discussion in Section
\ref{sec:Wsymm}.)

We stress that, due to the presence of $z$ in the coefficient of
the $d \xi$ terms on r.h.s., these equations \eqref{eq:exa13} are
\emph{not} in Ito form. On the other hand, they can be integrated.

\section{Discussion \& Conclusions}
\label{sec:discuss}

In a previous work \cite{GS17} we have classified admissible (on
physical and mathematical basis) transformations of Ito stochastic
differential equations, and hence types of possible symmetries of
these. This classifications yielded three types of symmetries,
i.e. standard deterministic symmetries, standard random ones, and
W-symmetries. The first two types have been studied, by ourselves
and different authors, in the literature
\cite{GGPR,GL1,GL2,GS17,Koz1,Koz2,Koz3,Koz2018,Koz18a,Koz18b,Lunini,SMS},
while W-symmetries had so far lacked attention.

In this paper, we have first reviewed relevant notions and results
in the recent literature devoted to symmetry of SDEs; in
particular we have stressed the relevance of the relation between
symmetries of an Ito equation and of the associated Stratonovich
one (actually, this is an equality for admissible symmetries), and
recalled how Kozlov theory makes use of symmetries to integrate --
at least partially, in which case we actually have a reduction --
Ito equations.

In the second and main part of the paper, we have studied
W-symmetries. In particular, in Sect.\ref{sec:Wsymm} we have
determined the class of vector fields qualifying as admissible
would-be W-symmetries, obtaining in particular that the action of
the $w$ variables must correspond to an origin-preserving action
of the linear conformal group -- thus essentially reduce to
dilations and/or rotations. In Sect \ref{sec:WSIS} we have
established the determining equations for W-symmetries of Ito and
Stratonovich SDEs, discussing the relation between their
solutions. This turns out to be less trivial than for standard
symmetries, and in particular symmetries acting as dilations in
the $w$ sector may not be shared by an Ito and the associated
Stratonovich equation, as also shown in concrete simple examples.

Finally, in Sect.\ref{sec:Wkoz} we have considered how
W-symmetries can be concretely used in studying Ito SDEs. We have
seen that the situation is different from the one which became
familiar with standard symmetries. In fact, once one has
determined a W-symmetry of a given Ito equation, one can reduce it
to a partially integrable equation by passing to symmetry-adapted
variables, but in general this produce a non-Ito equation. This
means in particular that the existing theory -- which only
considers Ito equations -- cannot be used for reduction under
multiple symmetries, and thus calls for extension of the theory to
a wider realm.

It should be recalled that more general (than Ito or Stratonovich)
types of stochastic differential equations do not have an equally
solid mathematical foundation; but they are nevertheless used in
several physical (and chemical) contexts \cite{Kampen}.

\newpage

\section*{Appendix A. The one-dimensional case}

{\small

In this Appendix we discuss the problem tackled in Section
\ref{sec:WSIS} in the simplified setting of scalar Ito equation.
This will allow to avoid plethora of indices and get a more clear
view of the reasoning and computations leading to our results
there.

It follows from our general discussion that in the scalar
(one-dimensional) case $$  d x \ = \ f (x,t) \, d t \ + \ \s (x,t)
\, d w \eqno(A.1) $$ the only nontrivial W-symmetries act on $w$
as dilations. This case is of course specially simple, and it is
worth looking at it specifically; we will use an obvious
simplified notation, and this will provide a check of our general
result in the simplest setting.

Now the Ito determining equations read $$ \vphi_t \ + \ f \,
\vphi_x \ - \ \vphi \, f_x \ + \ \frac12 \Delta \vphi  \ = \ 0 \ ,
\eqno(A.2) $$
 $$ \vphi_w \ + \ \s \, \vphi_x \ - \ \vphi \, \s_x \
- \ \s \, R \ = \ 0 \ . \eqno(A.3) $$

As for the Stratonovich determining equations for $$ d x \ = \ b
(x,t) \, d t \ + \ \s (x,t) \circ d w \ , \eqno(A.4) $$ these read
\begin{eqnarray*}
\vphi_t \ + \ b \, \vphi_x \ - \ \vphi \, b_x  &=& 0 \ , \\
\vphi_w \ + \ \s \, \vphi_x \ - \ \vphi \, \s_x \ - \ \s \, R &=&
0 \ . \end{eqnarray*}

The (common) second equation of these sets yields
$$ \vphi_w \ = \ \vphi \, \s_x \ - \ \s \, \vphi_x \ + \ \s \, R \
. $$ This in turn provides \begin{eqnarray*} \vphi_{xw} &=&
\vphi_x \, \s_x \ + \ \vphi \, \s_{xx} \ - \ \s_x \, \vphi_x \ - \
\s \, \vphi_{xx} \ + \ \s_x \, R \ , \\
\vphi_{ww} &=& \vphi_w \, \s_x \ - \ \s \, \vphi_{xw} \\
 &=& \vphi_w \, \s_x \ - \ \vphi \, \s \, \s_{xx} \ + \ \s^2 \,
 \vphi_{xx} \ - \ \s \, \s_x \, R \ . \end{eqnarray*}
With these, and some trivial algebra, we get $$ \Delta \vphi \ = \
\vphi \, \s_x^2 \ - \ \vphi_x \, \s \, \s_x \ + \ \vphi \, \s \,
\s_{xx} \ + \ 2 \, \s \, \s_x \, R \ . \eqno(A.5) $$

On the other hand, recalling that for the associated Stratonovich
equation $b = f - (1/2) \rho$, we readily get that the first
determining equation in the Stratonovich case reads $$ \vphi_t \ +
\ f \, \vphi_x \ - \ \vphi \, f_x \ + \ \frac12 \[ \vphi \, \s_x^2
\ + \ \vphi \, \s \, \s_{xx} \ - \ \vphi_x \, \s \s_x \]  \ = \  0
\ .
$$ This coincides with the first determining equation for the Ito
equation if and only if $$ \Delta \vphi \ = \ \vphi \, \s_x^2 \ +
\ \vphi \, \s \, \s_{xx} \ - \ \vphi_x \, \s \s_x \ . \eqno(A.6)
$$ As discussed above, it suffices that the equality holds when we
restrict to solutions to the second (common) equation in the sets
of determining equations.

Comparing (A.5) and (A.6) we see that the equations coincide, and
hence symmetries of the Ito and of the associated Stratonovich
equations also do, if and only if $$ \s \, \s_x \ R \ = \ 0 \ .
\eqno(A.7) $$ We do of course exclude the case $\s=0$ (or the
equation would not be a stochastic one), and also the case $R=0$
as in that case we have a standard symmetry (which is trivial as a
W-symmetry).

So in the end we have shown the following, which of course is a
special case of the general result (Theorem 1) obtained above:

\medskip\noindent
{\bf Lemma A.1.} {\it In the one dimensional case the nontrivial
W-symmetries of an Ito equation are shared by the associated
Stratonovich equation if and only if the diffusion coefficient $\s
(x,t)$ in eq.(A.1) satisfies $\s_x = 0$}.

}

\section*{Appendix B. Forbidden forms of W-symmetries}

{\small It turns out that W-symmetries can not take all forms,
i.e. some forms of the W-symmetry generator $X = \vphi^i (x,t;w)
\pa_i + (R^k_{\ m} w^m )\^\pa_k$ can not be realized.

In order to illustrate this, we will consider some situations
assuming a given shape for $X$ -- i.e. for the coefficients
$\vphi^i$ -- and showing there can be no Ito equation admitting
such W-symmetry.

We will just consider some one-dimensional cases and restrict to
the time-autonomous case (that is, $f$ and $\s$, and hence also
$\vphi$ are assumed to be independent of $t$), which will help
keeping computations simple and focus on the qualitative relevant
point.

At the moment we are not able to provide general results on which
shape of W-symmetries are possible or forbidden.

\medskip\noindent
{\bf Example B.1.} Let us make the ansatz
$$ \vphi (x,t;w) \ = \ p (x,t) \ e^w \ ; \eqno(B.1) $$
then differentiating the second determining equation (A.3) w.r.t.
$w$ we get
$$ e^w \ \[ p \ + \ \s \, p_x \ - \ p \, \s_x \] \ = \ 0 \ , $$
and hence
$$ \s (x,t) \ = \ h(t) \, p(x,t) \ + \ \int_0^x \frac{1}{p(y,t)}
\, d y \ . $$ When plugging this into (A.3) itself, we get
$$ R \ p(x,t) \ \[ h[t] \ + \ \int_0^x \frac{1}{p(y,t)}
\, d y \] \ = \ 0 \ . $$ But the only solutions to this is $R=0$,
in which case we have a standard symmetry (choosing $p(x,t) = 0$
gives a singular situation, and no symmetry anyway). \EOR

\medskip\noindent
{\bf Example B.2.} Let us now look for W-symmetry generators with
$$ \vphi (x,t,w) \ = \ p(x,t) \ w^2 \ ; \eqno(B.2) $$
in this case (A.3) reads
$$ 2 \, w \, p(x,t) \ - \ R \, \s (x,t) \ + \ w^2 \ \[ S[x,t] \,
p_x (x,t) \, - \, p(x,t) \, \s_x (x,t) \] \ = \ 0 \ . $$ The term
linear in $w$ enforces $p = 0$, hence there is no equation
admitting W-symmetries of the form (B.2). \EOR

\medskip\noindent
{\bf Example B.3.} Finally, we consider the separable ansatz $$
\vphi(x,t,w) \ = \ p(x) \ q(w) \ ; \eqno(B.3) $$ note here we must
assume $p\not= 0 \not= q$ to rule out trivial cases.

We know that there is at least one case admitting nontrivial
W-symmetries of this form, see Example 11 in the main text. Here
we show that is the \emph{only} class of
time-autonomous\footnote{A full discussion would also be possible
admitting time dependencies, but it would be too long to report
here.} scalar equations admitting W-symmetries of the form (B.3).

In fact, plugging the ansatz (B.3) into (A.3) and differentiating
w.r.t. $w$, we get
$$ p \, q'' \ + \ \s \, p_x  \, q' \ - \ \s_x \, p \, q' \ = \ 0 \
; $$ this also reads
$$ \frac{p \, \s_x \ - \ p_x \, \s}{p} \ = \ \frac{q''}{q'} \ . $$
As the l.h.s. only depends on $(x,t)$ and the r.h.s. only on $w$,
it must be
$$ \frac{p \, \s_x \ - \ p_x \, \s}{p} \ = \ K \ = \ \frac{q''}{q'}
\eqno(B.4) $$ for some constant $K$.

Let us first assume $K \not= 0$. Then the r.h.s. equality in (B.4)
yields immediately
$$ q(w) \ = \ c_1 \ e^{K w} \ + \ c_2  \eqno(B.5) $$ (note that we
must require $c_1 \not= 0$, or we would have a split symmetry);
while setting
$$ \s (x) \ = \ [c_3 \ + \ r(x)] \ p(x) \eqno(B.6) $$
the l.h.s. equality reads
$$ p (x) \ = \ - \frac{K}{r' (x)} \ . \eqno(B.7) $$
When we insert (B.5)--(B.7) into (A.3), and look at the
coefficient of $e^{K w}$ in this, we get
$$ c_1 \, K^3 \ = \ 0 \ . $$
Thus we are left only with non-viable options: $c_1 = 0$ would
give a split symmetry and $K=0$ was excluded by assumption.

So we are forced to assume $K=0$. Now the r.h.s. equality in (B.4)
yields
$$ q(w) \ = \ c_1 \, w \ + \ c_2 \eqno(B.8) $$ while the l.h.s.
one provides
$$ p(x) \ = \ c_3 \ \s (x,t) \ . \eqno(B.9) $$
These were obtained from a differential consequence of (A.3); when
we plug (B.8) and (B.9) into (A.3) itself, we obtain
$$ c_3 \ = \ R/c_1 \ . $$

We can now tackle (A.2), which is an expression linear in $w$ (all
the dependencies on $w$ are now explicit). Thus it splits into two
equations (corresponding to the vanishing of terms independent of
$w$ and of the coefficient of $w$); assuming $R \not= 0$ these
read
\begin{eqnarray*}
R [2 c_2 f \s_x + 2 \s (c_1 \s_x - c_2 f_x) + c_2 \s^2 \s_{xx}] &=& 0 \\
R [2 c_1 f \s_x - 2 c_1 \s f_x + c_1 \s^2 \s_{xx}] &=& 0 \ .
\end{eqnarray*}
Multiplying the first by $c_1$, the second by $c_2$ and taking the
difference, we get
$$ R \, c_1^2 \, \s \, \s_x \ = \ 0 \ . $$
As $c_1$, $R$ and $\s$ can not vanish, we must have $\s_x = 0$,
i.e. $\s (x) = \mu $ (with $\mu \not= 0$). With this, the two
equations reduce to
$$  c_2 \, \mu \ f_x  \ = \  0 \ = \ c_1 \, \mu \ f_x \ , $$
$$ f(x) \ = \ \la \ . $$

In conclusion, we have obtained
\begin{eqnarray*}
f(x) &=& \la \ , \\
\s (x) &=& \mu \ ; \\
\vphi &=& \mu \, R \ w \ + \ k_1 \, \mu \, R \ .
\end{eqnarray*}
Here $k_1 = (c_2 / c_1)$ is an arbitrary constant, so we actually
have two symmetry generators, i.e.
$$ X_1 \ = \ \mu \, w \, \pa_x \ + \ \pa_w \ , \ \ \
X_2 \ = \ k_1 \, \mu \, \pa_x \ + \ \pa_w \ ; $$ note however that
$X_2$ is a split W-symmetry. \EOR

}

\end{document}